\documentclass[10pt,twocolumn,letterpaper]{article}

\usepackage{iccv}              % To produce the CAMERA-READY version
\usepackage{multirow}
\definecolor{iccvblue}{rgb}{0.21,0.49,0.74}
\usepackage[pagebackref,breaklinks,colorlinks,allcolors=iccvblue]{hyperref}

\definecolor{pinklink}{RGB}{255,105,180} 
\def\OursDataset{MDD\xspace} 
\def\paperID{12712} % *** Enter the Paper ID here
\def\confName{ICCV}
\def\confYear{2025}

\title{MDD: A Dataset for Text-and-Music Conditioned Duet Dance Generation}

\author{
Prerit Gupta \quad
Jason Alexander Fotso-Puepi \quad
Zhengyuan Li \quad
Jay Mehta \quad
Aniket Bera \\
Purdue University, West Lafayette, IN, USA \\
{\tt\small \{gupta596, jfotsopu, li5280, mehta208, aniketbera\}@purdue.edu} \\
{\tt\small \href{https://gprerit96.github.io/mdd-page}{\texttt{\textcolor{pinklink}{https://gprerit96.github.io/mdd-page}}}}
\footnotetext{This paper has been accepted as an ICCV 2025 conference paper.}
}

\begin{document}
\twocolumn[{
\maketitle
\begin{center}
    \vspace{-20pt}
    {Accepted at ICCV 2025}
\end{center}
\begin{center}
  \centering
  \includegraphics[width=\textwidth]{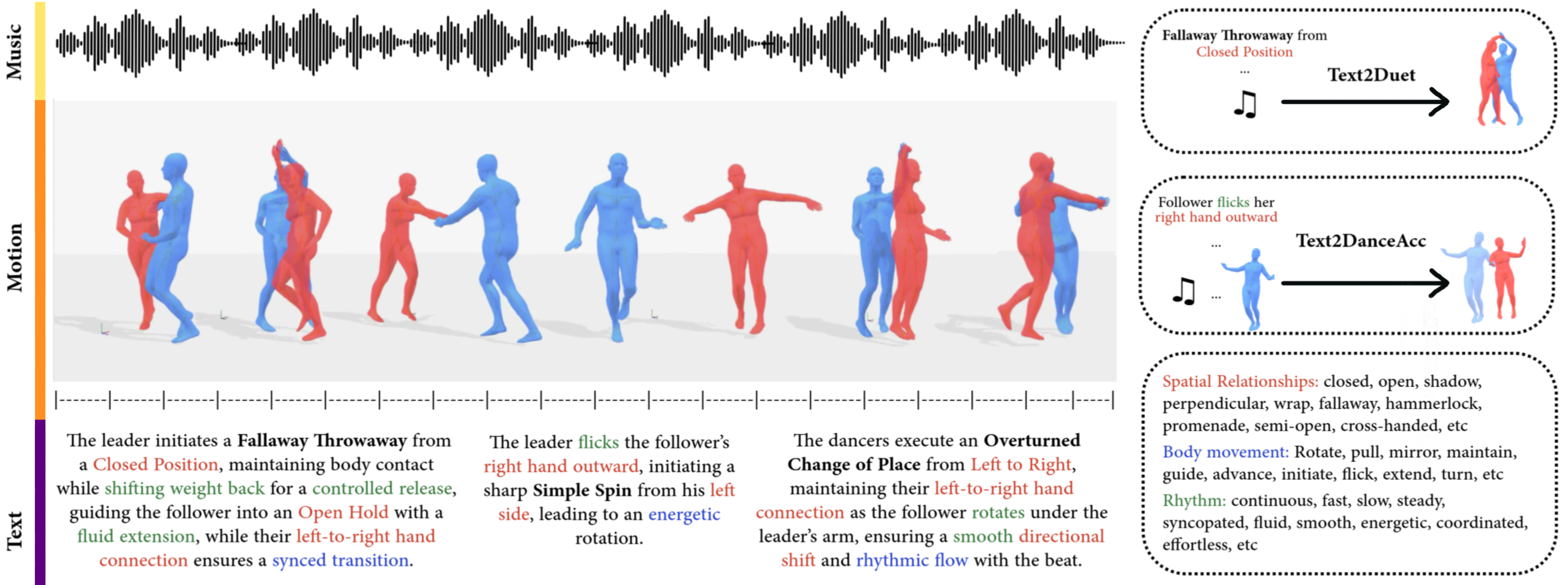}
  \captionof{figure}{Overview of the Multimodal DuetDance (MDD) dataset which includes duet dance motions, music and multi-faceted text annotations. Left: An example sample from the Jive genre with paired motion and text. Top right: The dataset enables two key downstream tasks—Text-to-Duet (Text2Duet) and Text-to-Dance Accompaniment (Text2DanceAcc). Bottom right: Examples of the diverse movement vocabulary captured in our annotations, highlighting spatial relationships, body dynamics, and rhythmic patterns.}
  \label{fig:cover}
\end{center}
}]

\begin{abstract}
We introduce Multimodal DuetDance (MDD), a diverse multimodal benchmark dataset designed for text-controlled and music-conditioned 3D duet dance motion generation. Our dataset comprises 620 minutes of high-quality motion capture data performed by professional dancers, synchronized with music, and detailed with over 10K fine-grained natural language descriptions. The annotations capture a rich movement vocabulary, detailing spatial relationships, body movements, and rhythm, making MDD the first dataset to seamlessly integrate human motions, music, and text for duet dance generation. We introduce two novel tasks supported by our dataset: (1) Text-to-Duet, where given music and a textual prompt, both the leader and follower dance motion are generated (2) Text-to-Dance Accompaniment, where given music, textual prompt, and the leader's motion, the follower's motion is generated in a cohesive, text-aligned manner. We include baseline evaluations on both tasks to support future research. Please refer to the project website for the latest updates.

% Our benchmark provides a robust foundation for text-controllable interactive two-person dance motion synthesis and text-driven reactive dance synthesis, fostering advancements in controllable interactive dance motion generation.
\end{abstract}

%  To advance these tasks, we introduce an Interactive Diffusion-based model conditioned on both music and text, incorporating Retrieval-Augmented Generation (RAG) to refine genre-specific motion synthesis.    
\section{Introduction}
\label{sec:intro}

%MUsic is considered by default when we talk about dance motion as any motion can be called as dance only when synchronized with music. So music is by default a condiition and I think, it might not be necessary here to mention why is needed as it's solely a dance generation task .

% \Zhengyuan{DEBUGGING}
Duet dancing is a complex form of interactive human motion that requires precise coordination and synchronization between two dancers. Unlike solo performances, duet dances involve intricate spatial relationships, dynamic partner interactions, and continuous mutual adaptation to music beats. The lead dancer directs the interaction through movement cues synchronized with the music beats, while the follower responds in real-time, maintaining harmony and adding stylistic variations. This interdependent relationship presents significant modeling challenges, including movement unpredictability, motion misalignments, and inconsistencies in interaction dynamics, making duet dance generation substantially more complex than single-person dance synthesis. To address these challenges, introducing explicit control mechanisms can significantly enhance the conditioning of the generation process, ensuring coherence and responsiveness in duet dance synthesis.

We emphasize the importance of control signals in generating duet dance motions. Generic motion generation often leverages natural language~\cite{zhang2024motiondiffuse, liang2024omg, Zhang_2023_CVPR, tevet2023human, meng2024rethinking, wang2023fg, athanasiou2024motionfix, yi2024generating}, while most dance generation works~\cite{siyao2023bailando++, li2021ai, tseng_edge_2022, ghosh2025duetgen} primarily rely on music due to its rhythmic and structural properties. The former line of research demonstrates that textual prompts specifying style, interaction, and timing enhance motion control and contextual awareness. Building on this, text-based control can improve duet dance generation by aligning movements with choreographic intent and preserving the intricate lead–follower dynamics. Well-crafted prompts enable better synchronization and genre-specific fidelity, thereby enhancing realism and coherence. Moreover, text input allows choreographers to define sequences, refine interaction styles, and enforce synchronization constraints more effectively.

Interactive motion generation has been explored in works such as InterGen~\cite{liang_intergen_2023} and Inter-X~\cite{xu_inter-x_2023}, which introduce large-scale datasets of interactive human motion with textual annotations, enabling two-person generation. However, these datasets primarily focus on general interactions and lack specialized professional movements (e.g., dance) and synchronized audio. Meanwhile, Duolando~\cite{siyao_duolando_2024} provides a duet dance benchmark aligned with music and motion, but it contains a limited number of samples and lacks textual annotations, making it difficult to generalize across diverse dance genres.

% MODIFIED
To address these limitations, we present \OursDataset as illustrated in \cref{fig:cover}, a large-scale professional-grade duet dance motion dataset with fine-grained text annotations that capture a rich movement vocabulary for spatial relationships, body movements, and rhythm. It contains over \textbf{4.4M} frames with at least \textbf{30} minutes of high-quality motion data for \textbf{15} different genres across Latin, Ballroom, and Social Dancing categories. The data was captured using Optitrack equipment with experienced dancers. We collected more than \textbf{10K} textual descriptions from annotators with diverse dance backgrounds, ranging from novices to dance experts. Our dataset supports two novel multi-modal tasks:
(1) \textbf{Text-to-Duet}: Conditional generation task that produces synchronized movements for both dancers given a text description and music. This task challenges models to generate plausible partner interactions based on text prompts while maintaining genre-specific styling and ensuring musical synchronization. 
(2) \textbf{Text-to-Dance Accompaniment}: Reactive generation task that synthesizes follower movements in response to the leader's actions, guided by textual prompts and musical context. These tasks represent fundamental challenges in multi-person motion synthesis and serve as benchmarks for evaluating models' ability to generate complex, interactive dance movements. 

In summary, our key contributions are:
\vspace{0.2cm}  
\begin{enumerate}
    \item We introduce \OursDataset, the first large-scale duet dance motion dataset spanning \textbf{15} different dance genres, featuring more than \textbf{10K} rich text descriptions across more than \textbf{10} hours of motion capture data.
    \item We provide fine-grained annotations that capture spatial relationships, body movement dynamics, and rhythmic elements between dance partners. %\Zhengyuan{could be merged with the first one}
    \item We introduce two new tasks that support our dataset with challenging multi-modal benchmarks: \textbf{Text-to-Duet} for generating coordinated partner movements from a text description and music, and \textbf{Text-to-Dance Accompaniment} for synthesizing context-aware follower motion given the leader's motion, music and text description.
    % \Zhengyuan{why are the two tasks meaningful? What's the difference between text2dance accompaniment and text-based reactive generation? What is the difference between two-person motion generation and Text2DuetDance? } 
\end{enumerate}

% \begin{equation}
%   E = m\cdot c^2
%   \label{eq:important}
% \end{equation}
% and
% \begin{equation}
%   v = a\cdot t.
%   \label{eq:also-important}
% \end{equation}

% \noindent
% Compare the following:\\
% \begin{tabular}{ll}
%  \verb'$conf_a$' &  $conf_a$ \\
%  \verb'$\mathit{conf}_a$' & $\mathit{conf}_a$
% \end{tabular}\\
% See The \TeX book, p165.

% \begin{figure*}
%   \centering
%   \begin{subfigure}{0.68\linewidth}
%     \fbox{\rule{0pt}{2in} \rule{.9\linewidth}{0pt}}
%     \caption{An example of a subfigure.}
%     \label{fig:short-a}
%   \end{subfigure}
%   \hfill
%   \begin{subfigure}{0.28\linewidth}
%     \fbox{\rule{0pt}{2in} \rule{.9\linewidth}{0pt}}
%     \caption{Another example of a subfigure.}
%     \label{fig:short-b}
%   \end{subfigure}
%   \caption{Example of a short caption, which should be centered.}
%   \label{fig:short}
% \end{figure*}

\section{Related Work}
\label{sec:related-work}
\subsection{Music-Dance Datasets}

% Several single-person dance motion datasets \cite{huang_dance_2021, alexanderson2023listen, luo2024popdg} have been developed, featuring music-aligned joint representations of motion captured through multi-view RGB videos and depth sensors. Among these, AIST++ \cite{li2021ai} emerged as one of the most widely used 3D choreography datasets, providing five hours of dance motion across ten different genres. Additionally, several datasets \cite{alexanderson2023listen, zhuang2022music2dance, li2023finedance} have leveraged high-quality motion capture technology, offering more precise movement representations. FineDance \cite{li2023finedance}, the largest single-person dance motion dataset, contains 14.6 hours of motion capture data spanning 22 different genres, making it a crucial resource for single-person dance motion generation.

% In the domain of duet dancing, only two datasets: DD100~\cite{siyao_duolando_2024} and InterDance~\cite{li2024interdance} have been developed to date. Siyao et al.~\cite{siyao_duolando_2024} introduced DD100, the first duet dance motion dataset, which includes 1.95 hours of motion capture data covering ten genres. More recently, InterDance~\cite{li2024interdance} expanded upon this by collecting 3.93 hours of duet dance motion across 15 genres. However, given the intricate interdependencies and nuanced partner interactions inherent in duet dancing, the current datasets remain insufficient in terms of sample diversity and generalizability across all genres.

Existing single-person dance datasets \cite{huang_dance_2021, alexanderson2023listen, luo2024popdg} feature music-aligned motion captured via RGB and depth sensors. AIST++ \cite{li2021ai} remains a widely used benchmark with five hours of 3D motion across ten genres. High-quality MoCap datasets like FineDance \cite{li2023finedance} offer greater precision, with 14.6 hours of motion spanning 22 genres—the largest for solo dance generation. In the domain of duet dancing, only two datasets: DD100~\cite{siyao_duolando_2024} and InterDance~\cite{li2024interdance} have been developed to date. Siyao et al.~\cite{siyao_duolando_2024} introduced DD100, the first duet dance motion dataset, which includes 1.95 hours of motion capture data covering ten genres. More recently, InterDance~\cite{li2024interdance} expanded upon this by collecting 3.93 hours of duet dance motion across 15 genres. However, given the intricate interdependencies and nuanced partner interactions inherent in duet dancing, the current datasets remain insufficient in terms of sample diversity and generalizability across all genres.

% In contrast, duet dance datasets are limited. DD100~\cite{siyao_duolando_2024} introduced 1.95 hours of partner motion across ten genres, while InterDance~\cite{li2024interdance} extended this to 3.93 hours over 15 genres. However, both fall short in capturing the full diversity and complexity of partner interactions across genres.

\subsection{Single-Person Dance Generation}
The study of single-person dance motion generation has been explored significantly in past few years \cite{Wang_2024_CVPR} 
\cite{Li_2024} \cite{zhuang2022music2dance}. Various state-of-the-art models \cite{huang_dance_2021} \cite{tseng_edge_2022} \cite{yang2023longdancediff} employ music-to-motion transformers, variational autoencoders or diffusion-based models to generate complex dance motions directly from music inputs. These methods often leverage large-scale datasets that capture various dance styles, allowing the models to learn the intricate connections between rhythm, body movement, and expressiveness. In contrast, duet dance motion generation remains largely underexplored due to the unavailability of a large-scale, high-quality dataset. Unlike single-person dance synthesis, duet dancing involves complex interactions between two dancers, including lead-follower dynamics, synchronized movements, and non-verbal communication. 

\subsection{Text-Conditioned Motion Generation}
Recent advances in text-to-motion generation~\cite{liang_intergen_2023, xu_inter-x_2023, petrovich_temos_2022, li2025simmotionedit, bhattacharya2021text2gestures, tevet2022motionclip} enable expressive human motion synthesis from natural language. TM2D~\cite{Gong2023TM2DBD} introduces a bimodal approach using both music and text, but suffers from distributional mismatch due to mixing motion data from dance-specific and generic datasets, limiting text controllability. Most existing work focuses on single-person dance generation, and thus face limitations to capture rich interpersonal dynamics inherent in duet dancing. Moreover, generic motion datasets lack vocabulary to represent the diversity of dance movements. To address this, we introduce \OursDataset, a benchmark for text-conditioned duet dance generation that enables studying partner interactions, coordination, and synchronization, moving beyond solo motion synthesis.

\subsection{Interactive Two-Person Generation}
Recent advances in multi-person motion generation have evolved from trajectory forecasting to full-body interaction synthesis. InterGen \cite{liang_intergen_2023} introduced the first large-scale two-person interaction motion dataset, while Inter-X \cite{xu_inter-x_2023} expanded this with fine-grained text annotations, enabling text-conditioned interactive motion generation. In duet dancing, Duolando \cite{siyao_duolando_2024} introduced the first Duet Dance Motion dataset with paired motion and music across 10 genres. However, its limited sample size (approximately 12 minutes per genre) and lack of structured multi-person motion generation tasks restrict its generalizability. To address these gaps, we introduce \OursDataset, the first dataset integrating text, motion, and music, enabling text-guided multi-person dance generation. We propose Text-to-Duet task, where duet dance motion is conditioned on both text prompts and music, introducing the first text-controlled duet dance generation task.

\subsection{Reactive Motion Generation}
Several prior works \cite{men_gan-based_2021, ghosh_remos_2023} have advanced the task of reactive motion synthesis, which focuses on generating one individual's motion in response to another’s actions. Recent works~\cite{xu2024regennet, cen2025ready} add text or constraints to guide generation more contextually. For duet dancing, Duolando~\cite{siyao_duolando_2024} adapts this by synthesizing the follower's motion based on the leader and music. However, this approach lacks fine-grained controllability, as the dataset provides only limited paired examples and lacks motion descriptions, making it difficult to generalize across different genres.  To address this, we introduce Text-to-Dance Accompaniment, extending reactive motion generation by incorporating rich textual annotations that capture spatial relationships, body movements and rhythm. By conditioning the follower’s motion on both the leader’s movements and textual descriptions, our approach enables more nuanced, expressive, and genre-adaptive duet dance synthesis.

\begin{figure*}[ht]
  \centering
  \includegraphics[width=\textwidth]{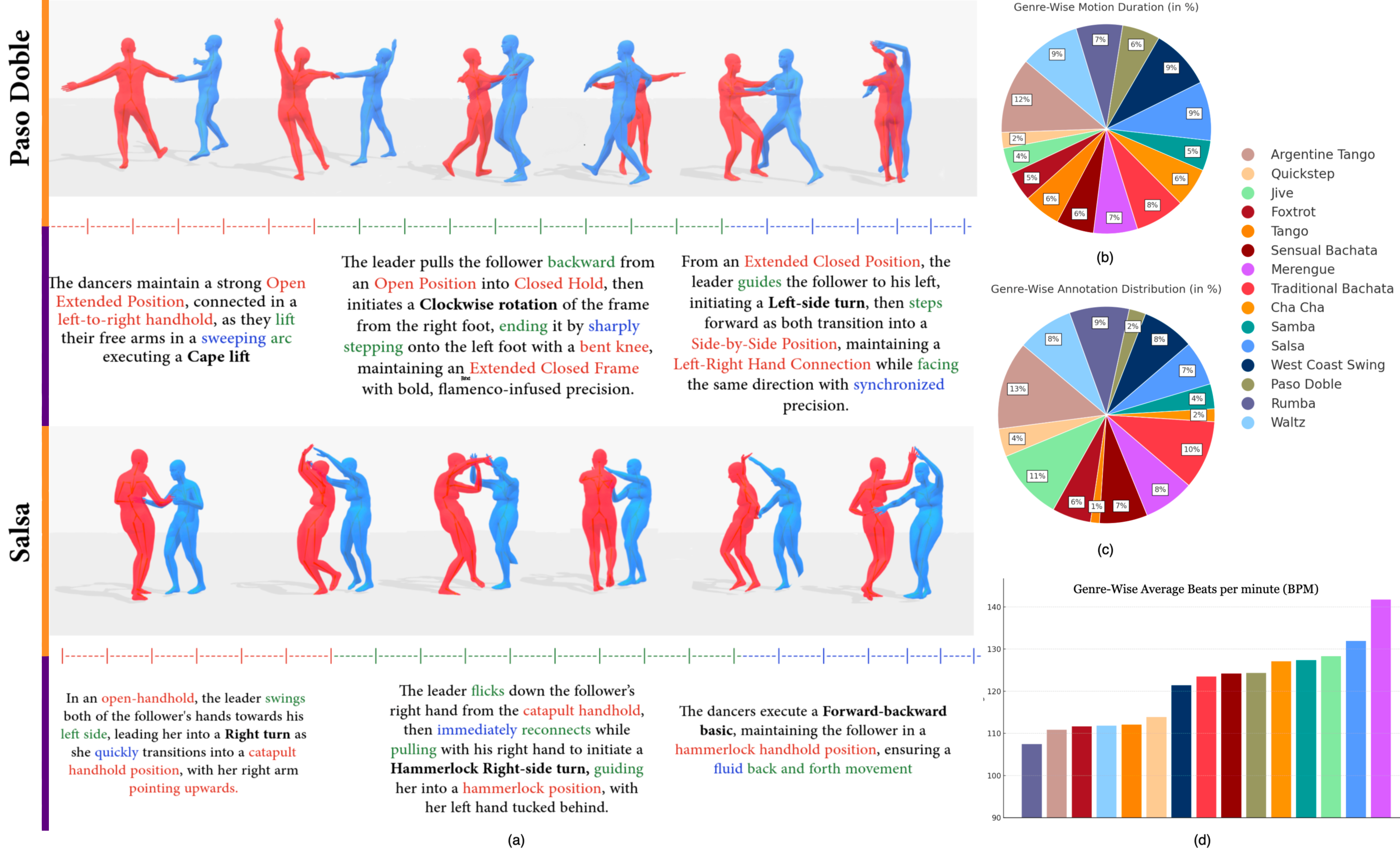}
  \caption{(a) We illustrate two samples from the MDD dataset with corresponding annotated descriptions. The first, from \textbf{Paso Doble}, features a \textit{Cape lift}, followed by an \textit{Open Spin} and a \textit{Left-side turn}. The second, from \textbf{Salsa}, shows a double-handed \textit{Right turn} leading into a \textit{Catapult handhold} followed by \textit{right hand flick} with a \textit{Right turn} ending into follower's \textit{Hammerlock position}. On the right, we present genre-wise statistics for (b) motion duration percentage, (c) annotation counts distribution, and (d) average Beats Per Minute (BPM)}
  \label{fig:genre-wise}
\end{figure*}

\section{\OursDataset Dataset}
\label{sec:dataset}

We present \OursDataset, a large-scale multimodal duet dance dataset consisting of 10.34 hours of motion capture data featuring more than 10K rich fine-grained text descriptions based on spatial relationship, body movements and rhythm. 

\subsection{Data Statistics}
Our dataset features high-fidelity motion capture data spanning over 15 diverse dance genres, categorized into three broad styles: (1) Ballroom (Waltz, Foxtrot, Quickstep, Tango), (2) Latin (Rumba, Jive, Cha Cha, Samba, Paso Doble) and (3) Social (Salsa, Traditional Bachata, Sensual Bachata, Merengue, West Coast Swing, Argentine Tango). The dataset was collected from 30 subjects (16 females, 14 males), ensuring a balanced gender representation while capturing a wide range of dance styles and individual interpretations. Each genre includes a minimum of 30 minutes of motion capture data, providing extensive coverage for model generalization for different movement patterns. \cref{fig:genre-wise} presents dataset insights with distribution statistics for different modalities: (a) Samples show high motion quality with rich annotations (b) Genre-wise motion duration distribution is relatively balanced (c) Annotation counts per genre are also mostly balanced, with slight variations for some genres having different duration length requirements per each annotated sample
 (d) Genres span a wide BPM range, from Rumba (lowest) to Merengue (highest). Please refer to the Supplementary Material for more details.

%Jay: Add important statistics & analysis here

\subsection{Data Collection}
The data collection process involved five main components: (1) Music selection (2) Motion capture (3) Motion post-processing  (4) Annotating Descriptions (5) Annotation Refining. Before starting the data collection process, an IRB approval was obtained from the university.

\subsubsection{Music Selection}
For music selection, priority was given to copyright-free sources. However, for certain genres, obtaining a sufficient number of copyright-free samples was challenging. In such cases, short excerpts (0.5 - 1 min) were randomly selected, ensuring compliance with fair use for research purposes. Each genre folder contained a diverse collection of approximately 50–60 unique music samples, providing dancers with a broad selection to choose from.

\begin{figure}[ht]
  \centering
  \includegraphics[width=\columnwidth]{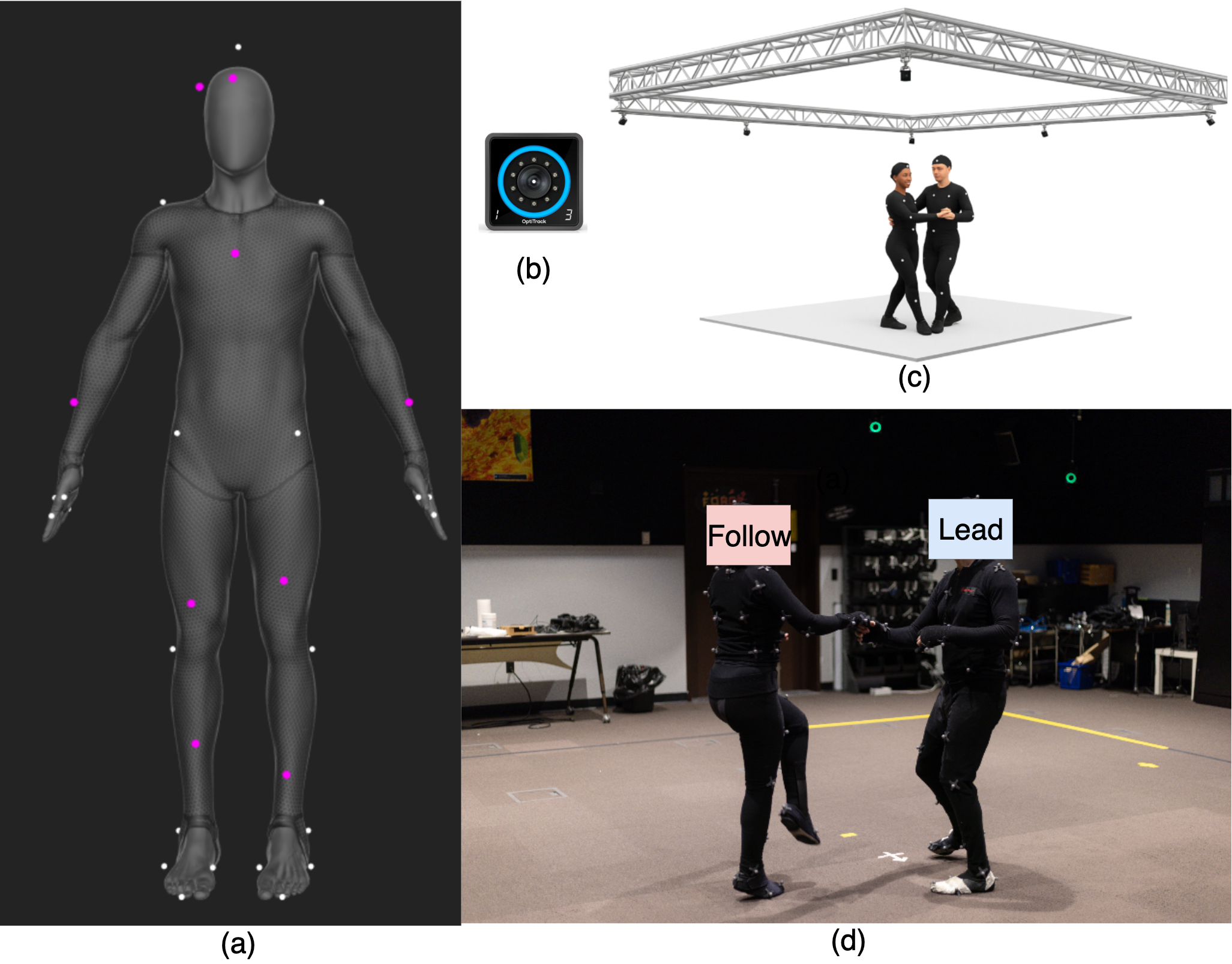}
  \caption{(a) Placement of 53 markers using the OptiTrack Motive software. (b) Infrared optical camera (Prime 17W) used for high-speed tracking at a resolution of 1280 × 1024. (c) Layout of the motion capture facility. (d) Snapshot of our data collection setup}
  \label{fig:mocap}
\end{figure}

\subsubsection{Motion Capture}
Motion capture was performed in a 24 × 24 ft. laboratory facility at 120 fps using an OptiTrack MoCap system \cite{optitrack_motive} with 16 infrared cameras optimally placed at varying heights (10–12 ft.). We used 53 retro-reflective markers optimally placed on the subjects who wore a form fitting suit as shown in \cref{fig:mocap} This configuration ensured precise data capture for natural, high-quality recordings to capture complex partner interactions.

We recruited 30 dancers, consisting of 16 females and 14 males, each with expertise in various partner-based dance styles across 15 distinct genres we selected. All participants were at an intermediate or advanced skill level, with a minimum of 3 years of experience in their respective genres. All participants came prepared with a diverse set of moves and they were allowed to freely adapt to the music, simulating a social dancing setting.

\subsubsection{Motion Postprocessing}

Recordings with marker swaps or major capture errors were discarded. For short-term occlusions occurring from partner contact, missing marker data was manually identified and interpolated using Spherical Interpolation\cite{1145217} from neighboring markers' data. We further performed post-processing to remove jitter, snapping, and contact noise through: (1) outlier removal via frame-wise distance thresholds and interpolation; (2) Gaussian filtering in low-motion regions to reduce noise while preserving dynamics; (3) correction of zero-pose vectors using linear interpolation; and (4) block-aware blending for smooth transitions, especially in hand and arm movements. We observed that these steps enhanced motion realism and continuity in partner interactions.

We use SMPL-X \cite{SMPL-X:2019} parametric model for motion representation due to its ability to represent different human body poses. Specifically, the SMPL-X parameters are defined by the body pose parameters $\theta \in \mathbb{R}^{N \times 55 \times 3}$, shape parameters $\beta \in \mathbb{R}^{N \times 10}$ and the translation parameters $t \in \mathbb{R}^{N \times 3}$, where $N$ is the number of the frames. Each sample was converted to this SMPL-X format by training a parametric model  based on the optimization algorithm described in Inter-X \cite{xu_inter-x_2023} where we also learn the shape parameters $\beta$ by minimizing the following objective function:

\begin{equation}
E(\theta,t,\beta) = \lambda_1\frac{1}{N}\sum_{j \in \mathcal{J}} \lambda_p||\textbf{J}_j(M(\theta,t,\beta)) - g_j||_2^2 + \lambda_2||\theta||_2^2 
\end{equation}

\noindent where $M$ denotes the SMPL-X parameters, $\mathcal{J}$ denotes the joints set, $J_j$ is the joint regressor function for joint $j$, $g$ is the skeleton captured key points and $\lambda_1, \lambda_2$ and $\lambda_p$ are different weights.

%  A Gaussian motion processing filter was applied to smoothen out the jitter in the samples described as follows:

% \[
% \hat{y}_i = \sum_{j=-m}^{m} c_j y_{i+j}
% \]

% where $\hat{y}_i$ is the smoothed value at index $i$, $y_{i+j}$ are the original signal values within the window centered at $i$, $c_j$ are the convolution coefficients  and $m$ is the window half-size, so the total window size is $2m + 1$ for the filter. 

\begin{figure}[ht]
  \centering
  \includegraphics[width=\columnwidth]{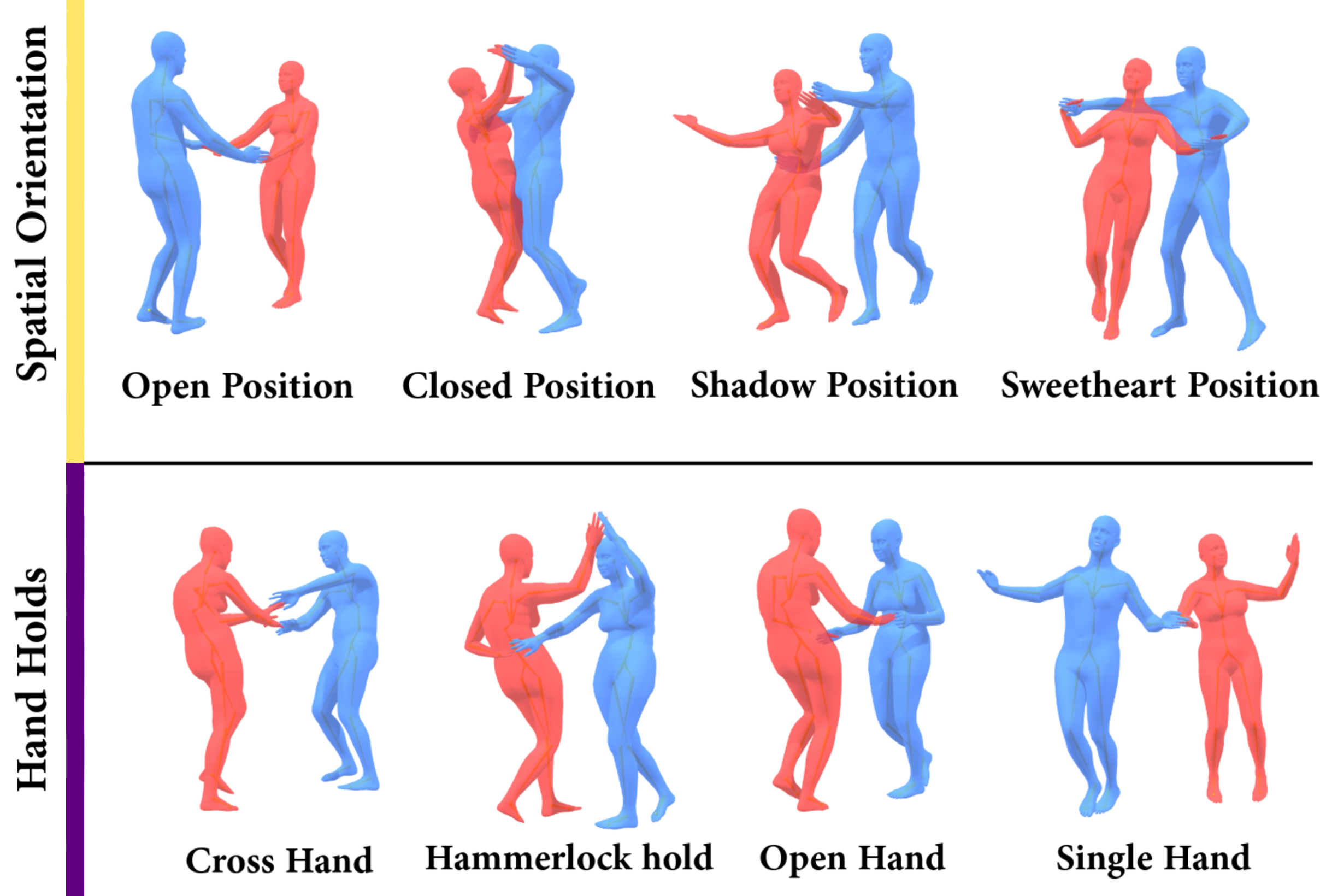}
  \caption{Some example snapshots showing different spatial orientations and handholds captured in our fine-grained text annotations}
  \vspace{3mm}
  \label{fig:positions}
\end{figure}

% \begin{table}[ht]
%   \centering
%   \begin{tabular}{@{}p{1.5cm}p{1,5cm}p{5cm}@{}}
%     \toprule
%     Category & Type & Example values \\
%     \midrule
%      & Interaction Position & Open, Closed, Semi-open, Shadow, Side-by-side, Promenade, Ballroom hold, Counter-balance, Counter-weight \\
%     Spatial Relationship & Orientation & Facing each other, Back-to-back, Side-to-side, Offset (e.g., right or left), Diagonal \\
%     & Connection Points & Hand-to-hand, Hand-to-waist, Hand-to-shoulder, Arm hook, Elbow hold, Waist hold, Shoulder hold, Leg wrap, Foot connection \\
%     \midrule
%      & Movement Type & Lifting, Twisting, Rotating, Bending, Extending, Contracting, Isolating, Waving, Pushing, Pulling \\
%     Body Movement & Body Parts & Head, Neck, Shoulders, Arms, Elbows, Hands, Waist, Hips, Knees, Legs, Feet, Toes \\
%     \midrule
%      & Energy & Soft/gentle, Medium, Sharp/intense, Explosive, Fluid, Staccato \\
%     Rhythm & Tempo & Fast, Slow, Syncopated, Polyrhythmic, Continuous, Pulsed \\
%     \bottomrule
%   \end{tabular}
%   \caption{Description of pointers used to create fine-grained text descriptions. We provide the cheatsheet to annotators to facilitate high-quality annotations.}
%   \label{tab:annotation-set}
% \end{table}

\begin{table}[ht]
  \scriptsize
  \renewcommand{\arraystretch}{0.95}
  \setlength{\tabcolsep}{3pt}
  \centering
  \begin{tabular}{@{}p{1.3cm}p{2cm}p{4.5cm}@{}}
    \toprule
    \textbf{Category} & \textbf{Type} & \textbf{Example Values} \\
    \midrule
    \multirow{3}{*}{\shortstack[l]{Spatial \\ Relationship}} 
      & Interaction Position 
      & Open, Closed, Semi-open, Shadow, Side-by-side, Promenade, Ballroom hold, Counter-balance, Counter-weight \\
    & Orientation 
      & Facing each other, Back-to-back, Side-to-side, Offset, Diagonal \\
    & Connection Points 
      & Hand-to-hand, Hand-to-waist, Hand-to-shoulder, Arm hook, Elbow hold, Waist hold, Shoulder hold, Leg wrap \\
    \midrule
    \multirow{2}{*}{\shortstack[l]{Body \\ Movement}} 
      & Movement Type 
      & Lifting, Twisting, Rotating, Bending, Extending, Contracting, Isolating, Pushing, Pulling \\
    & Body Parts 
      & Head, Neck, Shoulders, Arms, Elbows, Hands, Waist, Hips, Knees, Legs, Feet, Toes \\
    \midrule
    \multirow{2}{*}{Rhythm} 
      & Energy 
      & Soft/gentle, Medium, Fluid, Sharp/intense, Explosive, Staccato \\
    & Tempo 
      & Fast, Slow, Syncopated, Polyrhythmic, Continuous, Pulsed \\
    \bottomrule
  \end{tabular}
  \caption{Examples of categories and attributes used to guide fine-grained text annotations. Annotators referred to this structured list for consistency.}
  \label{tab:annotation-set}
\end{table}

\subsubsection{Annotating Descriptions}

To collect detailed annotations, we developed a user-friendly web-based annotation tool that presented annotators with the rendered versions of the motions using ait-viewer\cite{Kaufmann_Vechev_aitviewer_2022}. The camera parameters were dynamically adjusted to continuously track the dancers, ensuring an optimal viewing perspective for annotators. Annotators were instructed to segment the videos into discrete clips that best preserved the semantic integrity of each interaction. The annotation tool enforced a maximum segment duration of 10 seconds, adhering to the same guidelines as the InterHuman dataset \cite{liang_intergen_2023}. 

To maintain annotation quality, we asked the same participant dancers to provide textual descriptions corresponding to their motion capture data. Annotators were provided with detailed guideline documents to ensure consistency in motion terminology. Each description was required to incorporate key vocabulary from three predefined categories: (1) Spatial Relationships, (2) Body Movements, and (3) Rhythm, as outlined in the guidelines. Some of the examples belonging to this set are outlined in \cref{tab:annotation-set}. \cref{fig:positions} shows some of the different spatial relationships captured in our dataset. Please refer to the supplementary materials for more details.

\subsubsection{Annotation Refining}
To improve clarity and consistency, we used GPT-4o~\cite{islam2024gpt} to refine annotation grammar while preserving their original semantic meaning. The refined annotations were then reviewed by a second set of annotators with dancing expertise (specifically, the dance partners of the original annotators) to ensure accuracy and correctness. The final annotations average 41 words, longer than existing motion-text datasets, highlighting their fine-grained detail. This multi-stage process enhances linguistic quality and enriches the movement vocabulary to better represent each dance sequence.

\begin{table*}[ht]
  \centering
  % After removing 2 columns and adding 1, we now have 10 columns total
  \begin{tabular}{@{}p{3cm}p{1.2cm}p{0.8cm}p{0.8cm}p{0.9cm}p{1cm}p{1cm}p{1cm}p{1cm}p{1.6cm}@{}}
    \toprule
    Dataset & Interactive & Music & Text & Genres & Subjects & Frames & Mocap & $T$ & Descriptions \\
    \midrule
    DanceRevolution \cite{huang_dance_2021} &  & \checkmark &  & 3 &  & 648K &  & 12h & - \\
    DanceNet \cite{zhuang2022music2dance}   &  & \checkmark &  & 2 & - & 207K & \checkmark & 0.96h & - \\
    AIST++ \cite{li2021ai}                  &  & \checkmark &  & 10 & 30 & - &  & 5.2h & - \\
    Motorica Dance \cite{alexanderson2023listen} &  & \checkmark &  & 8 & - & 1.3M & \checkmark & 6.22h & - \\ 
    FineDance \cite{li2023finedance}        &  & \checkmark &  & 22 & 27 & 10K & \checkmark & 14.6h & - \\
    PopDanceSet \cite{luo2024popdg}         &  & \checkmark &  & 19 & 132 & 769K &  & 3.56h & - \\
    \midrule
    InterHuman \cite{liang_intergen_2023}   & \checkmark &  & \checkmark & 13 & 89 & 1.7M &  & 6.56h & 23,337 \\
    Inter-X \cite{xu_inter-x_2023}          & \checkmark &  & \checkmark & - & - & 8.1M & \checkmark & 18.75h & 34,164 \\
    \midrule
    AIOZ-GDANCE \cite{le2023music}          & \checkmark & \checkmark &  & 7 & 4000+ & 1.8M &  & 16.7h & - \\
    DD100 \cite{siyao_duolando_2024}        & \checkmark & \checkmark &  & 10 & 10 & 210K & \checkmark & 1.95h & - \\
    ReMoS \cite{ghosh_remos_2023}           & \checkmark & \checkmark &  & 2 & 9 & 275.7K &  & 2.04h & - \\
    InterDance \cite{li2024interdance}      & \checkmark & \checkmark &  & 15 & - & 1.7M & \checkmark & 3.93h & - \\
    \midrule
    \OursDataset (Ours)                        & \checkmark & \checkmark & \checkmark & 15 & 30 & 4.4M & \checkmark & 10.34h & 10,187 \\
    \bottomrule
  \end{tabular}
  \caption{Comparison between \OursDataset and other related datasets. \OursDataset is a large-scale two-person dance dataset that provides both music and text annotations. It can be seen that our dataset contains a high diversity of genres , total number of frames and comparable number of descriptions with other Interactive datasets }
  \label{tab:data-comp}
\end{table*}

\subsection{Comparison to Related Datasets}

We compare \OursDataset with existing human motion datasets, particularly those focusing on interactions and dance movements, as shown in \cref{tab:data-comp}. Our analysis shows that \OursDataset stands out with these distinct features:
\textbf{(1) Multi-modal:} Existing two-person interactive motion datasets are based on either integrating music and motion \cite{li2024interdance}\cite{siyao_duolando_2024} or text and motion \cite{liang_intergen_2023} \cite{xu_inter-x_2023}. To the best of our knowledge, \OursDataset is the first dataset to comprehensively integrate all three modalities: motion, music, and text descriptions. It is capable to support motion generation tasks conditioned for both music and text, specifically for duet dancing. 
\textbf{(2) Largest Duet Dance Dataset:} With 10.34 hours of high-quality motion capture data spanning across 15 different dance genres, \OursDataset offers substantial scale being the largest duet dance dataset that is helpful for a good model generalization for every movement style.
\textbf{(3) Fine-grained annotations:} While Intergen \cite{liang_intergen_2023} contains only generic descriptions and Inter-X \cite{xu_inter-x_2023} contains detailed descriptions based on generic human body movement, our collected textual description annotations are meticulously designed to capture dance-specific movement using technical dance terminology, incorporating spatial relationships between dance partners, style-specific movement vocabulary and rhythmic elements.

\section{Tasks}
\label{sec:tasks}

In this section, we first discuss our duet interaction representation, and then we introduce two novel tasks that leverage our \OursDataset dataset: \textit{Text-to-Duet} and \textit{Text-to-Dance Accompaniment}

% Let the text space be  , music space be, The motion space $\mathcal{M}$ is defined as $\mathcal{M} \in $. A sample $j$ is defined in the space
% $s_j$, $s_j in (\mathcal{T}, \mathcal{M}, \mathcal{X})$
% For a given text prompt $t \in \mathcal{T}$ where $\mathcal{T}$ is the description space of text prompts, input music feature  $m \in \mathcal{M}$ where $\mathcal{M} \in \mathbb{R}^{T \times d}$ where $d$ is the dimension of music feature , lead's motion $x_l$ where $x_l \in \mathcal{M}$ where   and $x_f \in \mathcal{M}$. A sample in our dataset is defined in the space 

% A sample $s_j$ in our dataset is defined in the joint space:
% \begin{equation}
%     s_j \in (\mathcal{T}, \mathcal{M}, \mathcal{X}).
% \end{equation}

% For a given text prompt $t \in \mathcal{T}$, where $\mathcal{T}$ represents the space of textual descriptions, we define the input music feature $m \in \mathcal{M}$, where:
% \begin{equation}
%     \mathcal{M} \subset \mathbb{R}^{T \times d},
% \end{equation}
% where $T$ is the temporal length and $d$ is the feature dimension of the music representation.

% The motion sequences consist of:
% - The **lead's motion** $x_l \in \mathcal{X}$
% - The **follower's motion** $x_f \in \mathcal{X}$

% Thus, a complete sample in our dataset consists of a tuple:
% \begin{equation}
%     s_j = (t, m, x_l, x_f),
% \end{equation}
% where each component belongs to its respective space.

\subsection{Duet Interaction Representation}

A duet motion interaction $\mathbf{x} \in \mathcal{X} \times \mathcal{X}$, is defined as a collection of the leader's motion $\mathbf{x_l} = \{x_l^i\}_{i=1}^{N}$ and the follower's motion $\mathbf{x_f} = \{x_f^i\}_{i=1}^{N}$ where all interactive pairs $x^j = (x_l^j, x_f^j) \in \mathbf{x}$ are naturally synchronized. Here, the motion space $\mathcal{X}$ is defined as $\mathcal{X} \subset \mathbb{R}^{N \times J \times 3}$ where $N$ is the number of frames and $J$ is the number of joints represented.

We define the music space $\mathcal{M}$ as $\mathcal{M} \subset \mathbb{R}^{N \times d_m}$ where $d_m$ is the feature dimension of the music representation and the text descriptive space by $\mathcal{C}$. For a given text description $c \in \mathcal{C}$, music feature $m \in \mathcal{M}$, our multi-modal duet interaction sample in our dataset is defined as $s = (c, m, \mathbf{x})$. Here, $s \in \mathcal{S}$ where the sample joint space $\mathcal{S}$ defined as $\mathcal{S} = (\mathcal{C} \times \mathcal{M} \times \mathcal{X} \times \mathcal{X})$

\subsection{Text-to-Duet}
Given a natural language description $c \in \mathcal{C}$ and music $m \in \mathcal{M}$, the task for Text-conditioned Duet Dance Generation aims to generate plausible duet interactions $(\mathbf{x_l},\mathbf{x_f})$ that are synchronous with music and semantically align with the text description $c$ . Formally, the task is to learn a function\textbf{ $F: F(c,m) \mapsto \mathbf{x}$}. The task can be regarded as an extension of text-guided two-person motion generation~\cite{liang_intergen_2023} in the dance scenario.

% \begin{equation}
%     F(c,m) \mapsto \mathbf{x}
% \end{equation}

% The task can be regarded as an extension of text-guided two-person motion generation~\cite{liang_intergen_2023} in the dance scenario.
% where $F: \mathcal{C} \times \mathcal{M} \rightarrow \mathcal{X} \times \mathcal{X}$

% where $t \in \mathcal{T}$ is the input text description, $m \in \mathcal{M}$ is the input music features, $\mathcal{X}_L$ represents the motion space for the lead dancer and $\mathcal{X}_F$ represents the motion space for the follow dancer

\subsection{Text-to-Dance Accompaniment}

Given a natural language description $c \in \mathcal{C}$, music $m \in \mathcal{M}$ and leader's motion $\mathbf{x_l} \in \mathcal{X}$, the task for Text-conditioned Dance Accompaniment aims to generate the follower's motion $\mathbf{x_f} \in \mathcal{X}$ such that the duet interactions $(\mathbf{x_l},\mathbf{x_f})$ are interactively coherent, synchronous with music and semantically align with the text description. Formally, the task is to learn a function $G: G(c,m, \mathbf{x_l}) \mapsto \mathbf{x_f}$. This task can be viewed as an extension of human action-reaction synthesis~\cite{xu2024regennet} tailored to dance scenarios or, alternatively, as an extension of dance accompaniment~\cite{siyao_duolando_2024} guided by text.

% \begin{equation}
%     G(c,m, \mathbf{x_l}) \mapsto \mathbf{x_f}
% \end{equation}

% This task can be viewed as an extension of human action-reaction synthesis~\cite{xu2024regennet} tailored to dance scenarios or, alternatively, as an extension of dance accompaniment~\cite{siyao_duolando_2024} guided by text.

% where $G: \mathcal{T} \times \mathcal{M} \times \mathcal{X} \rightarrow \mathcal{X}$

% For a given time step $t$, the follower's motion is generated as:
% \begin{equation}
% x_F^t = g(t, m, x_L^{1:t}, x_F^{1:t-1})
% \end{equation}

% where $x_L^{1:t}$ represents the lead dancer's motion history up to time $t$, $x_F^{1:t-1}$ represents the follower's own motion history, $t$ is the text description and $m$ represents the music features

\section{Experiments}
\label{sec:experiments}
In this section, we present evaluation metrics, implementation details, and report both quantitative and qualitative results to assess the performance  of different baselines.

\subsection{Evaluation Metrics}
For evaluating models for the proposed tasks, we adapt metrics from previous text-to-motion~\cite{liang_intergen_2023, guo2022generating} and music-to-motion~\cite{siyao_duolando_2024, siyao2022bailando} works. For text-related metrics, we train the evaluator as in InterGen~\cite{liang_intergen_2023} on our dataset.

\begin{enumerate}
    \item \emph{Frechet Inception Distance (FID)}. Evaluates the distributional similarity between the ground truth and generated motions, irrespective of the input conditions. 
    \item \emph{Multimodal Distance (MM Dist)}. Measures the alignment between text conditions and generated motions based on feature-space distances.
    \item \emph{R-Precision}. Measures the alignment between the text conditions and the motions through retrieval accuracies within a batch. 
    \item \emph{Diversity}. Quantifies the variation across generated motions regardless of input conditions.
    \item \emph{Multimodality (MModality)}. Evaluates the diversity of generated motions conditioned on same textual input.
    \item \emph{Beat Echo Degree (BED)}. Measures temporal  synchronization between the leader's and follower's motions.
    \item \emph{Beat-Align Score (BAS)}. Assesses synchronization of each dancer's motion with the underlying music beats.
\end{enumerate}

% \subsection{Implementation Details}
% Since no previous methods can be directly applied to our task, we modify MDM~\cite{tevet2023human} and InterGen~\cite{liang_intergen_2023} for Text-to-Duet task. For MDM, we keep the diffusion process unchanged and concatenate the music features to the input sequence. For InterGen, we use cross attention in each block to inject music features. Unless specified, we follow~\cite{siyao_duolando_2024} and extract mel frequency cepstral coefficients (MFCC) features with librosa~\cite{mcfee2015librosa}. We additionally experiment with different InterGen variants with partial conditions or different music feature representations~\cite {dhariwal2020jukebox}. All the models are trained for $3000$ epochs with the AdamW~\cite{loshchilov2017decoupled} optimizer. The evaluation of R-Precision is done with a batch size of $64$. 

% For Text-to-Dance Accompaniment, we extend Duolando \cite{siyao_duolando_2024} by integrating a CLIP-based text encoder as input to GPT, incorporated via cross-modal attention to predict the follower’s motion. As in the previous section, our model utilizes MFCC-based music features, while other settings remain consistent with Duolando.

\subsection{Implementation Details}

Since no existing methods are directly applicable to our task, we adapt MDM~\cite{tevet2023human} and InterGen~\cite{liang_intergen_2023} for Text-to-Duet task, and Duolando~\cite{siyao_duolando_2024} for Text-to-Dance Accompaniment task. For MDM, we retain the original diffusion process and concatenate music features with the input motion. For InterGen, we introduce cross-attention layers in each transformer block to inject music features effectively. For Duolando, we extend the original model by incorporating CLIP-based text encodings as input to the GPT-style generator via cross-modal attention, enabling the generation of follower's motion based on textual context as well. Unless otherwise specified, we follow Duolando~\cite{siyao_duolando_2024} and extract 54-dimensional music representation based on mel-frequency cepstral coefficients (MFCCs) using Librosa~\cite{mcfee2015librosa}. For InterGen (both), we additionally experiment with music features extracted from Jukebox encoder~\cite{dhariwal2020jukebox}, which yields 4800-dimensional representations derived from a large 5B-parameter pre-trained model. All models are trained for 3000 epochs using AdamW optimizer~\cite{loshchilov2017decoupled} with a batch size of $64$.

\subsection{Quantitative Results}

Each adapted baseline is trained and evaluated in three configurations: (1) text-only (no music), (2) music-only (no text), and (3) both (text + music).

\paragraph{Text-to-Duet} As shown in \cref{tab:text2duet}, InterGen consistently outperforms MDM, highlighting its suitability for interactive generation. Among InterGen variants, the text-only model achieves the best performance, reflecting strong alignment with textual input. Using Jukebox~\cite{dhariwal2020jukebox} embeddings yields slight improvements over MFCCs, suggesting that richer music representations enhance generation. MDM exhibits a high BAS score, likely due to frequent jittery motions that the metric tends to reward. Meanwhile, Duolando~\cite{siyao_duolando_2024} reports the BAS of ground-truth data lower than that of the generated motion, underscoring the need to interpret this metric with caution. In contrast, BED correlates more consistently with fidelity-based metrics like FID.

\paragraph{Text-to-Dance Accompaniment} As shown in \cref{tab:text2reactive}, the multimodal model outperforms both text-only and music-only variants, confirming the advantage of conditioning generation on both modalities. The music-only model also outperforms the text-only variant in most metrics, likely due to the model architecture being originally designed to integrate motion and music features, which are more naturally aligned in shape and semantics than text.

% For Text-to-Dance Accompaniment, we extend Duolando \cite{siyao_duolando_2024} by integrating a CLIP-based text encoder as input to GPT, incorporated via cross-modal attention to predict the follower’s motion. As in the previous section, our model utilizes MFCC-based music features, while other settings remain consistent with Duolando. 

% We train two versions of the model: (1) Music-only model - trained using only music and paired motion (2) Multimodal model – trained using both music and text along with paired motions.

\begin{figure}[ht]
  \centering
  \includegraphics[width=\linewidth]{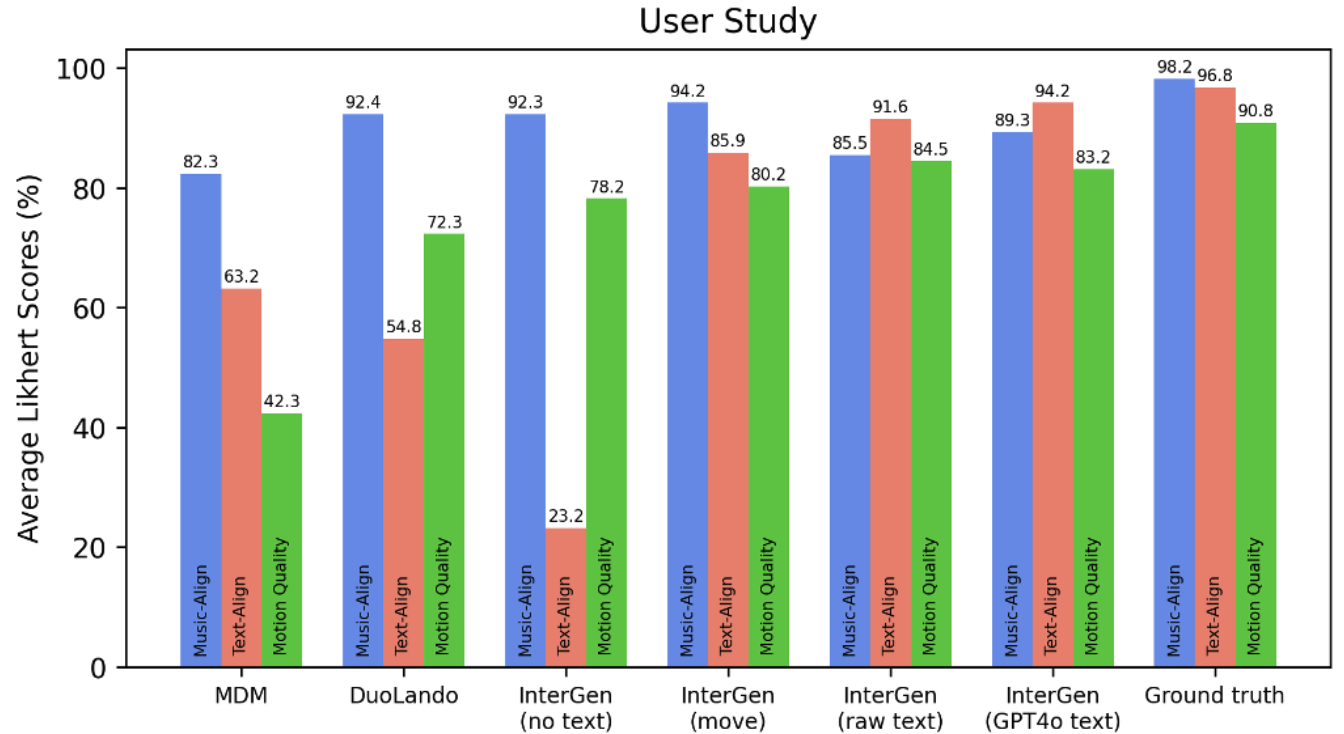}
  \caption{User Study Results}
  \label{fig:user-study}
\end{figure}

\begin{figure*}[ht]
  \centering
  \includegraphics[width=\linewidth]{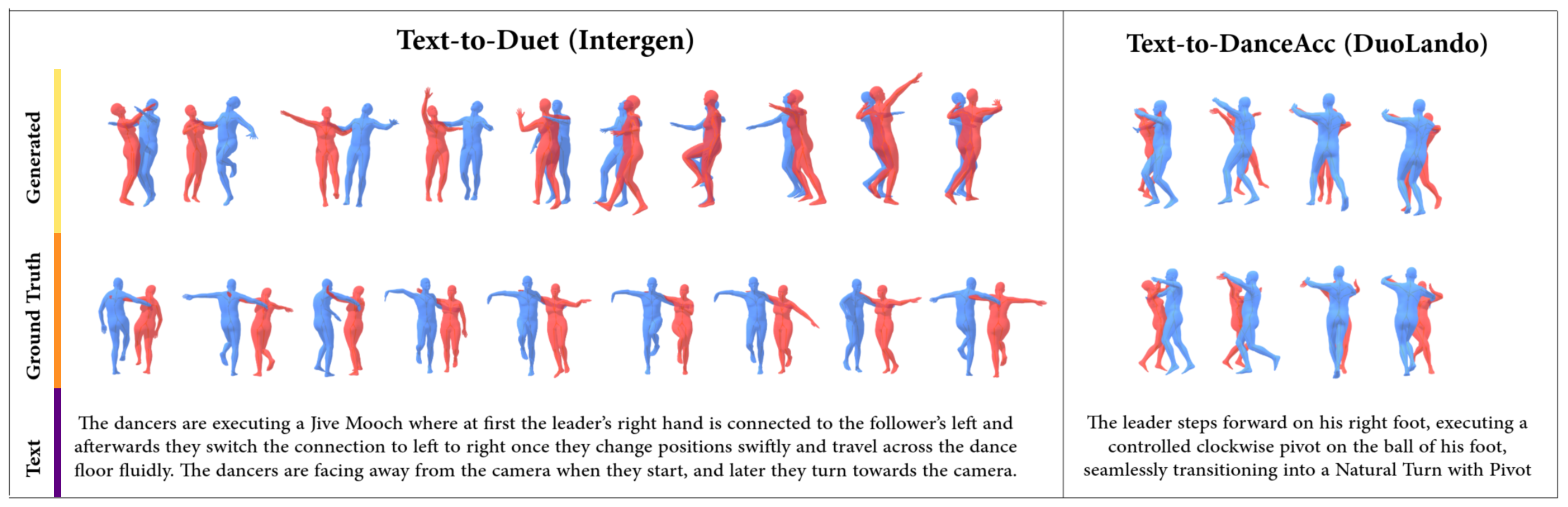}
  \caption{Qualitative comparison of generated samples using InterGen for the Text-to-Duet task (left, Jive sample) and Duolando for the Text-to-Dance Accompaniment task (right, Waltz sample) against ground truth. Both models are trained with text and music modalities.}
  \label{fig:result}
  \vspace{2mm}
\end{figure*}

\begin{table*}[h]
  \centering
  \resizebox{\textwidth}{!}{\begin{tabular}{lccccccccc}
    % 

    % \textbf{Methods} & \textbf{R-Precision (top 1)}& \textbf{R-Precision (top 2)} & \textbf{R-Precision (top 3)} & \textbf{FID} & \textbf{MM Dist} & \textbf{Diversity} & \textbf{MModality} & \textbf{BED} & \textbf{BAS} \\
    % \midrule
    \toprule
        \textbf{Methods} 
          & \multicolumn{3}{c}{\textbf{R-Precision}$\uparrow$} 
          & \textbf{FID}$\downarrow$
          & \textbf{MM Dist}$\downarrow$
          & \textbf{Diversity}$\rightarrow$
          & \textbf{MModality}$\uparrow$
          & \textbf{BED} $\uparrow$
          & \textbf{BAS}$\uparrow$ \\
        \cmidrule(lr){2-4}
        & \textbf{Top 1} & \textbf{Top 2} & \textbf{Top 3}
        & & & & & & \\
        \midrule
    Ground Truth & 0.231& 0.3984& 0.5219 & 0.065& 0.0770& 1.387& - & 0.327 & 0.170 \\
    \midrule
    MDM (text-only) & 0.082 & 0.124 & 0.192 & 1.420 & 2.133 & 1.216 & 0.811 & 0.211 & 0.186 \\
    MDM (music-only) & 0.041 & 0.102 & 0.135 & 2.241 & 2.471 & 1.192 & 0.411 & 0.210 & 0.192 \\
    MDM (both) & 0.061& 0.108& 0.163& 1.739& 2.244& 1.235& 0.787& 0.194& \textbf{0.231} \\
    \midrule
    InterGen (text-only) & {0.113}& {0.223}& {0.305}& \textbf{0.405}& {1.462}& 1.405& 1.231& 0.422 &0.194\\
    InterGen (music-only) & 0.023 & 0.067 & 0.088 & 2.014& 2.526 & 1.300& \textbf{1.768}& 0.364 & 0.163\\
    InterGen (both) & 0.105 & 0.206& 0.302& 0.426& 1.532 & {1.380}& 1.352& 0.385& 0.185\\
    InterGen \textit{w. Jukebox} (both) &  \textbf{0.138}& \textbf{0.245}& \textbf{0.341}& 0.410& \textbf{1.396}& \textbf{1.388}& 1.330& \textbf{0.454}&0.184\\
    \bottomrule
  \end{tabular}}
  \vspace{1mm}
  \caption{Quantitative evaluation of adapted baseline models for Text-to-Duet task. \textbf{Bold} indicates the best result.}
  \vspace{2mm}
  \label{tab:text2duet}
\end{table*}

\begin{table*}[!h]
  \centering
  \setlength{\tabcolsep}{8pt} % Adjust column spacing
  \begin{tabular}{l ccc c c c c c}
    \toprule
    \textbf{Methods}
      & \multicolumn{3}{c}{\textbf{R-Precision}$\uparrow$}
      & \textbf{FID}$\downarrow$
      & \textbf{MM Dist}$\downarrow$
      & \textbf{Diversity}$\rightarrow$
      & \textbf{BED}$\uparrow$
      & \textbf{BAS}$\uparrow$ \\
    \cmidrule(lr){2-4}
    & \textbf{Top 1} & \textbf{Top 2} & \textbf{Top 3}
    & & & & & \\
    \midrule
        Ground Truth & 0.231& 0.3984& 0.522 & 0.065& 0.0770& 1.387& 0.327 & 0.170 \\
    \midrule
    DuoLando (text-only) & 0.047 & 0.121 & 0.182 & 1.538  & 2.811 & 1.422 & 0.311 & 0.195 \\
    DuoLando (music-only)     & 0.069&0.141 &0.202 &0.721 & 2.633& \textbf{1.390}& 0.305 & 0.216 \\
    DuoLando (both)     & \textbf{0.078} &\textbf{0.156} &\textbf{0.219} & \textbf{0.698} & \textbf{2.113}& 1.371& \textbf{0.395} & \textbf{0.224} \\
    \bottomrule
  \end{tabular}
  \vspace{1mm}
  \caption{Quantitative evaluation of adapted baseline models for Text-to-Dance Accompaniment task. \textbf{Bold} indicates the best result.}
  \label{tab:text2reactive}
\end{table*}

\subsection{Qualitative Results}
We conducted a user study ~\cref{fig:user-study} with 20 participants to qualitatively evaluate different baseline models and assess text ablation. Each participant was shown 21 music-rendered videos along with text prompts (3 per experiment). For each sample, participants were asked to answer three key questions: \textit{(1) Which motion better aligns semantically with the textual description? (2) Which motion is better synchronized with the musical beats? (3) Which motion has higher overall quality (e.g., naturalness, smoothness etc)?} Users rated motion generated from the GPT-4o-refined text (InterGen) higher than both the raw text and move-name versions, highlighting the value of LLM-based refinement and preferring them over other baselines. We also visualize generated output of InterGen (for Text-to-Duet) and DuoLando (for Text-to-Dance Accompaniment) and compare them with ground truth sample in ~\cref{fig:result}.

\section{Conclusion}
\label{sec:conclusion}
We present a large-scale comprehensive dataset that provides a robust foundation for advancing research in text-driven multi-agent dance generation. Spanning 620 minutes of motion capture data across 15 dance genres and over 10,000 detailed text descriptions, \OursDataset surpasses existing datasets in both scale and annotation depth. Its integration of paired motion, music alignment, and fine-grained text enables new research in interactive animation, automated dance generation, and human interaction modeling. We also introduce two benchmark tasks: Text-to-Duet and Text-to-Dance Accompaniment.
\section*{Acknowledgment}
We would like to thank the Envision Center at Purdue University for providing access to their motion capture laboratory, which was instrumental in collecting the dataset used in this work. We are also grateful to the dancers and volunteers who participated in the motion capture sessions or contributed to the data collection process in any way. 
\label{sec:conclusion}
{
    \small
    \bibliographystyle{ieeenat_fullname}
    \bibliography{main}

\begin{thebibliography}{44}
\providecommand{\natexlab}[1]{#1}
\providecommand{\url}[1]{\texttt{#1}}
\expandafter\ifx\csname urlstyle\endcsname\relax
  \providecommand{\doi}[1]{doi: #1}\else
  \providecommand{\doi}{doi: \begingroup \urlstyle{rm}\Url}\fi

\bibitem[Alexanderson et~al.(2023)Alexanderson, Nagy, Beskow, and Henter]{alexanderson2023listen}
Simon Alexanderson, Rajmund Nagy, Jonas Beskow, and Gustav~Eje Henter.
\newblock Listen, denoise, action! audio-driven motion synthesis with diffusion models.
\newblock \emph{ACM Transactions on Graphics (TOG)}, 42\penalty0 (4):\penalty0 1--20, 2023.

\bibitem[Athanasiou et~al.(2024)Athanasiou, Cseke, Diomataris, Black, and Varol]{athanasiou2024motionfix}
Nikos Athanasiou, Alp{\'a}r Cseke, Markos Diomataris, Michael~J Black, and G{\"u}l Varol.
\newblock Motionfix: Text-driven 3d human motion editing.
\newblock In \emph{SIGGRAPH Asia 2024 Conference Papers}, pages 1--11, 2024.

\bibitem[Bhattacharya et~al.(2021)Bhattacharya, Rewkowski, Banerjee, Guhan, Bera, and Manocha]{bhattacharya2021text2gestures}
Uttaran Bhattacharya, Nicholas Rewkowski, Abhishek Banerjee, Pooja Guhan, Aniket Bera, and Dinesh Manocha.
\newblock Text2gestures: A transformer-based network for generating emotive body gestures for virtual agents.
\newblock In \emph{2021 IEEE virtual reality and 3D user interfaces (VR)}, pages 1--10. IEEE, 2021.

\bibitem[Cen et~al.(2025)Cen, Pi, Peng, Shuai, Shen, Bao, Zhou, and Hu]{cen2025ready}
Zhi Cen, Huaijin Pi, Sida Peng, Qing Shuai, Yujun Shen, Hujun Bao, Xiaowei Zhou, and Ruizhen Hu.
\newblock Ready-to-react: Online reaction policy for two-character interaction generation.
\newblock \emph{arXiv preprint arXiv:2502.20370}, 2025.

\bibitem[Dhariwal et~al.(2020)Dhariwal, Jun, Payne, Kim, Radford, and Sutskever]{dhariwal2020jukebox}
Prafulla Dhariwal, Heewoo Jun, Christine Payne, Jong~Wook Kim, Alec Radford, and Ilya Sutskever.
\newblock Jukebox: A generative model for music.
\newblock \emph{arXiv preprint arXiv:2005.00341}, 2020.

\bibitem[Ghosh et~al.(2024)Ghosh, Dabral, Golyanik, Theobalt, and Slusallek]{ghosh_remos_2023}
Anindita Ghosh, Rishabh Dabral, Vladislav Golyanik, Christian Theobalt, and Philipp Slusallek.
\newblock Remos: 3d motion-conditioned reaction synthesis for two-person interactions.
\newblock In \emph{European Conference on Computer Vision}, pages 418--437. Springer, 2024.

\bibitem[Ghosh et~al.(2025)Ghosh, Zhou, Dabral, Wang, Golyanik, Theobalt, Slusallek, and Guo]{ghosh2025duetgen}
Anindita Ghosh, Bing Zhou, Rishabh Dabral, Jian Wang, Vladislav Golyanik, Christian Theobalt, Philipp Slusallek, and Chuan Guo.
\newblock Duetgen: Music driven two-person dance generation via hierarchical masked modeling.
\newblock In \emph{Proceedings of the Special Interest Group on Computer Graphics and Interactive Techniques Conference Conference Papers}, pages 1--11, 2025.

\bibitem[Gong et~al.(2023)Gong, Lian, Chang, Guo, Jiang, Zuo, Mi, and Wang]{Gong2023TM2DBD}
Kehong Gong, Dongze Lian, Heng Chang, Chuan Guo, Zihang Jiang, Xinxin Zuo, Michael~Bi Mi, and Xinchao Wang.
\newblock Tm2d: Bimodality driven 3d dance generation via music-text integration.
\newblock In \emph{Proceedings of the IEEE/CVF International Conference on Computer Vision}, pages 9942--9952, 2023.

\bibitem[Guo et~al.(2022)Guo, Zou, Zuo, Wang, Ji, Li, and Cheng]{guo2022generating}
Chuan Guo, Shihao Zou, Xinxin Zuo, Sen Wang, Wei Ji, Xingyu Li, and Li Cheng.
\newblock Generating diverse and natural 3d human motions from text.
\newblock In \emph{Proceedings of the IEEE/CVF conference on computer vision and pattern recognition}, pages 5152--5161, 2022.

\bibitem[Huang et~al.(2020)Huang, Hu, Wu, Sawada, Zhang, and Jiang]{huang_dance_2021}
Ruozi Huang, Huang Hu, Wei Wu, Kei Sawada, Mi Zhang, and Daxin Jiang.
\newblock Dance revolution: Long-term dance generation with music via curriculum learning.
\newblock In \emph{International conference on learning representations}, 2020.

\bibitem[Islam and Moushi(2024)]{islam2024gpt}
Raisa Islam and Owana~Marzia Moushi.
\newblock Gpt-4o: The cutting-edge advancement in multimodal llm.
\newblock \emph{Authorea Preprints}, 2024.

\bibitem[Kaufmann et~al.(2022)Kaufmann, Vechev, and Mylonopoulos]{Kaufmann_Vechev_aitviewer_2022}
Manuel Kaufmann, Velko Vechev, and Dario Mylonopoulos.
\newblock {aitviewer}, 2022.

\bibitem[Le et~al.(2023)Le, Pham, Do, Tjiputra, Tran, and Nguyen]{le2023music}
Nhat Le, Thang Pham, Tuong Do, Erman Tjiputra, Quang~D Tran, and Anh Nguyen.
\newblock Music-driven group choreography.
\newblock In \emph{Proceedings of the IEEE/CVF Conference on Computer Vision and Pattern Recognition}, pages 8673--8682, 2023.

\bibitem[Li et~al.(2021)Li, Yang, Ross, and Kanazawa]{li2021ai}
Ruilong Li, Shan Yang, David~A Ross, and Angjoo Kanazawa.
\newblock Ai choreographer: Music conditioned 3d dance generation with aist++.
\newblock In \emph{Proceedings of the IEEE/CVF international conference on computer vision}, pages 13401--13412, 2021.

\bibitem[Li et~al.(2023)Li, Zhao, Zhang, Su, Ren, Zhang, Tang, and Li]{li2023finedance}
Ronghui Li, Junfan Zhao, Yachao Zhang, Mingyang Su, Zeping Ren, Han Zhang, Yansong Tang, and Xiu Li.
\newblock Finedance: A fine-grained choreography dataset for 3d full body dance generation.
\newblock In \emph{Proceedings of the IEEE/CVF International Conference on Computer Vision}, pages 10234--10243, 2023.

\bibitem[Li et~al.(2024{\natexlab{a}})Li, Dai, Zhang, Li, Yang, Guo, and Li]{Li_2024}
Ronghui Li, Yuqin Dai, Yachao Zhang, Jun Li, Jian Yang, Jie Guo, and Xiu Li.
\newblock Exploring multi-modal control in music-driven dance generation.
\newblock In \emph{ICASSP 2024-2024 IEEE International Conference on Acoustics, Speech and Signal Processing (ICASSP)}, pages 8281--8285. IEEE, 2024{\natexlab{a}}.

\bibitem[Li et~al.(2024{\natexlab{b}})Li, Zhang, Zhang, Zhang, Su, Guo, Liu, Liu, and Li]{li2024interdance}
Ronghui Li, Youliang Zhang, Yachao Zhang, Yuxiang Zhang, Mingyang Su, Jie Guo, Ziwei Liu, Yebin Liu, and Xiu Li.
\newblock Interdance: Reactive 3d dance generation with realistic duet interactions.
\newblock \emph{arXiv preprint arXiv:2412.16982}, 2024{\natexlab{b}}.

\bibitem[Li et~al.(2025)Li, Cheng, Ghosh, Bhattacharya, Gui, and Bera]{li2025simmotionedit}
Zhengyuan Li, Kai Cheng, Anindita Ghosh, Uttaran Bhattacharya, Liangyan Gui, and Aniket Bera.
\newblock Simmotionedit: Text-based human motion editing with motion similarity prediction.
\newblock In \emph{Proceedings of the Computer Vision and Pattern Recognition Conference}, pages 27827--27837, 2025.

\bibitem[Liang et~al.(2024{\natexlab{a}})Liang, Bao, Zhang, Ren, Xu, Yang, Chen, Yu, and Xu]{liang2024omg}
Han Liang, Jiacheng Bao, Ruichi Zhang, Sihan Ren, Yuecheng Xu, Sibei Yang, Xin Chen, Jingyi Yu, and Lan Xu.
\newblock Omg: Towards open-vocabulary motion generation via mixture of controllers.
\newblock In \emph{Proceedings of the IEEE/CVF Conference on Computer Vision and Pattern Recognition}, pages 482--493, 2024{\natexlab{a}}.

\bibitem[Liang et~al.(2024{\natexlab{b}})Liang, Zhang, Li, Yu, and Xu]{liang_intergen_2023}
Han Liang, Wenqian Zhang, Wenxuan Li, Jingyi Yu, and Lan Xu.
\newblock Intergen: Diffusion-based multi-human motion generation under complex interactions.
\newblock \emph{International Journal of Computer Vision}, 132\penalty0 (9):\penalty0 3463--3483, 2024{\natexlab{b}}.

\bibitem[Loshchilov and Hutter(2017)]{loshchilov2017decoupled}
Ilya Loshchilov and Frank Hutter.
\newblock Decoupled weight decay regularization.
\newblock \emph{arXiv preprint arXiv:1711.05101}, 2017.

\bibitem[Luo et~al.(2024)Luo, Ren, Hu, Huang, and Yao]{luo2024popdg}
Zhenye Luo, Min Ren, Xuecai Hu, Yongzhen Huang, and Li Yao.
\newblock Popdg: Popular 3d dance generation with popdanceset.
\newblock In \emph{Proceedings of the IEEE/CVF Conference on Computer Vision and Pattern Recognition}, pages 26984--26993, 2024.

\bibitem[McFee et~al.(2015)McFee, Raffel, Liang, Ellis, McVicar, Battenberg, and Nieto]{mcfee2015librosa}
Brian McFee, Colin Raffel, Dawen Liang, Daniel~PW Ellis, Matt McVicar, Eric Battenberg, and Oriol Nieto.
\newblock librosa: Audio and music signal analysis in python.
\newblock \emph{SciPy}, 2015:\penalty0 18--24, 2015.

\bibitem[Men et~al.(2022)Men, Shum, Ho, and Leung]{men_gan-based_2021}
Qianhui Men, Hubert~PH Shum, Edmond~SL Ho, and Howard Leung.
\newblock Gan-based reactive motion synthesis with class-aware discriminators for human--human interaction.
\newblock \emph{Computers \& Graphics}, 102:\penalty0 634--645, 2022.

\bibitem[Meng et~al.(2024)Meng, Xie, Peng, Han, and Jiang]{meng2024rethinking}
Zichong Meng, Yiming Xie, Xiaogang Peng, Zeyu Han, and Huaizu Jiang.
\newblock Rethinking diffusion for text-driven human motion generation.
\newblock \emph{arXiv preprint arXiv:2411.16575}, 2024.

\bibitem[{NaturalPoint, Inc.}(2019)]{optitrack_motive}
{NaturalPoint, Inc.}
\newblock {OptiTrack Motive: Motion Capture Software}.
\newblock \url{https://optitrack.com/products/motive/}, 2019.
\newblock Version 2.0 or later.

\bibitem[Pavlakos et~al.(2019)Pavlakos, Choutas, Ghorbani, Bolkart, Osman, Tzionas, and Black]{SMPL-X:2019}
Georgios Pavlakos, Vasileios Choutas, Nima Ghorbani, Timo Bolkart, Ahmed A.~A. Osman, Dimitrios Tzionas, and Michael~J. Black.
\newblock Expressive body capture: {3D} hands, face, and body from a single image.
\newblock In \emph{Proceedings IEEE Conf. on Computer Vision and Pattern Recognition (CVPR)}, pages 10975--10985, 2019.

\bibitem[Petrovich et~al.(2022)Petrovich, Black, and Varol]{petrovich_temos_2022}
Mathis Petrovich, Michael~J Black, and G{\"u}l Varol.
\newblock Temos: Generating diverse human motions from textual descriptions.
\newblock In \emph{European Conference on Computer Vision}, pages 480--497. Springer, 2022.

\bibitem[Siyao et~al.(2022)Siyao, Yu, Gu, Lin, Wang, Qian, Loy, and Liu]{siyao2022bailando}
Li Siyao, Weijiang Yu, Tianpei Gu, Chunze Lin, Quan Wang, Chen Qian, Chen~Change Loy, and Ziwei Liu.
\newblock Bailando: 3d dance generation by actor-critic gpt with choreographic memory.
\newblock In \emph{Proceedings of the IEEE/CVF Conference on Computer Vision and Pattern Recognition}, pages 11050--11059, 2022.

\bibitem[Siyao et~al.(2023)Siyao, Yu, Gu, Lin, Wang, Qian, Loy, and Liu]{siyao2023bailando++}
Li Siyao, Weijiang Yu, Tianpei Gu, Chunze Lin, Quan Wang, Chen Qian, Chen~Change Loy, and Ziwei Liu.
\newblock Bailando++: 3d dance gpt with choreographic memory.
\newblock \emph{IEEE Transactions on Pattern Analysis and Machine Intelligence}, 45\penalty0 (12):\penalty0 14192--14207, 2023.

\bibitem[Siyao et~al.(2024)Siyao, Gu, Yang, Lin, Liu, Ding, Yang, and Loy]{siyao_duolando_2024}
Li Siyao, Tianpei Gu, Zhitao Yang, Zhengyu Lin, Ziwei Liu, Henghui Ding, Lei Yang, and Chen~Change Loy.
\newblock Duolando: Follower gpt with off-policy reinforcement learning for dance accompaniment.
\newblock \emph{arXiv preprint arXiv:2403.18811}, 2024.

\bibitem[Smith and Abel(1987)]{1145217}
J. Smith and J. Abel.
\newblock The spherical interpolation method of source localization.
\newblock \emph{IEEE Journal of Oceanic Engineering}, 12\penalty0 (1):\penalty0 246--252, 1987.

\bibitem[Tevet et~al.(2022)Tevet, Gordon, Hertz, Bermano, and Cohen-Or]{tevet2022motionclip}
Guy Tevet, Brian Gordon, Amir Hertz, Amit~H Bermano, and Daniel Cohen-Or.
\newblock Motionclip: Exposing human motion generation to clip space.
\newblock In \emph{European Conference on Computer Vision}, pages 358--374. Springer, 2022.

\bibitem[Tevet et~al.(2023)Tevet, Raab, Gordon, Shafir, Cohen-or, and Bermano]{tevet2023human}
Guy Tevet, Sigal Raab, Brian Gordon, Yoni Shafir, Daniel Cohen-or, and Amit~Haim Bermano.
\newblock Human motion diffusion model.
\newblock In \emph{The Eleventh International Conference on Learning Representations}, 2023.

\bibitem[Tseng et~al.(2023)Tseng, Castellon, and Liu]{tseng_edge_2022}
Jonathan Tseng, Rodrigo Castellon, and Karen Liu.
\newblock Edge: Editable dance generation from music.
\newblock In \emph{Proceedings of the IEEE/CVF Conference on Computer Vision and Pattern Recognition}, pages 448--458, 2023.

\bibitem[Wang et~al.(2024)Wang, Li, Lin, Zhai, Lin, Yang, Zhang, Liu, and Wang]{Wang_2024_CVPR}
Tan Wang, Linjie Li, Kevin Lin, Yuanhao Zhai, Chung-Ching Lin, Zhengyuan Yang, Hanwang Zhang, Zicheng Liu, and Lijuan Wang.
\newblock Disco: Disentangled control for realistic human dance generation.
\newblock In \emph{Proceedings of the IEEE/CVF Conference on Computer Vision and Pattern Recognition (CVPR)}, pages 9326--9336, 2024.

\bibitem[Wang et~al.(2023)Wang, Leng, Li, Wu, and Liang]{wang2023fg}
Yin Wang, Zhiying Leng, Frederick~WB Li, Shun-Cheng Wu, and Xiaohui Liang.
\newblock Fg-t2m: Fine-grained text-driven human motion generation via diffusion model.
\newblock In \emph{Proceedings of the IEEE/CVF International Conference on Computer Vision}, pages 22035--22044, 2023.

\bibitem[Xu et~al.(2024{\natexlab{a}})Xu, Lv, Yan, Jin, Wu, Xu, Liu, Zhou, Rao, Sheng, et~al.]{xu_inter-x_2023}
Liang Xu, Xintao Lv, Yichao Yan, Xin Jin, Shuwen Wu, Congsheng Xu, Yifan Liu, Yizhou Zhou, Fengyun Rao, Xingdong Sheng, et~al.
\newblock Inter-x: Towards versatile human-human interaction analysis.
\newblock In \emph{Proceedings of the IEEE/CVF Conference on Computer Vision and Pattern Recognition}, pages 22260--22271, 2024{\natexlab{a}}.

\bibitem[Xu et~al.(2024{\natexlab{b}})Xu, Zhou, Yan, Jin, Zhu, Rao, Yang, and Zeng]{xu2024regennet}
Liang Xu, Yizhou Zhou, Yichao Yan, Xin Jin, Wenhan Zhu, Fengyun Rao, Xiaokang Yang, and Wenjun Zeng.
\newblock Regennet: Towards human action-reaction synthesis.
\newblock In \emph{Proceedings of the IEEE/CVF Conference on Computer Vision and Pattern Recognition}, pages 1759--1769, 2024{\natexlab{b}}.

\bibitem[Yang et~al.(2023)Yang, Yang, and Wang]{yang2023longdancediff}
Siqi Yang, Zejun Yang, and Zhisheng Wang.
\newblock Longdancediff: Long-term dance generation with conditional diffusion model.
\newblock \emph{arXiv preprint arXiv:2308.11945}, 2023.

\bibitem[Yi et~al.(2024)Yi, Thies, Black, Peng, and Rempe]{yi2024generating}
Hongwei Yi, Justus Thies, Michael~J Black, Xue~Bin Peng, and Davis Rempe.
\newblock Generating human interaction motions in scenes with text control.
\newblock In \emph{European Conference on Computer Vision}, pages 246--263. Springer, 2024.

\bibitem[Zhang et~al.(2023)Zhang, Zhang, Cun, Zhang, Zhao, Lu, Shen, and Shan]{Zhang_2023_CVPR}
Jianrong Zhang, Yangsong Zhang, Xiaodong Cun, Yong Zhang, Hongwei Zhao, Hongtao Lu, Xi Shen, and Ying Shan.
\newblock Generating human motion from textual descriptions with discrete representations.
\newblock In \emph{Proceedings of the IEEE/CVF Conference on Computer Vision and Pattern Recognition (CVPR)}, pages 14730--14740, 2023.

\bibitem[Zhang et~al.(2024)Zhang, Cai, Pan, Hong, Guo, Yang, and Liu]{zhang2024motiondiffuse}
Mingyuan Zhang, Zhongang Cai, Liang Pan, Fangzhou Hong, Xinying Guo, Lei Yang, and Ziwei Liu.
\newblock Motiondiffuse: Text-driven human motion generation with diffusion model.
\newblock \emph{IEEE transactions on pattern analysis and machine intelligence}, 46\penalty0 (6):\penalty0 4115--4128, 2024.

\bibitem[Zhuang et~al.(2022)Zhuang, Wang, Chai, Wang, Shao, and Xia]{zhuang2022music2dance}
Wenlin Zhuang, Congyi Wang, Jinxiang Chai, Yangang Wang, Ming Shao, and Siyu Xia.
\newblock Music2dance: Dancenet for music-driven dance generation.
\newblock \emph{ACM Transactions on Multimedia Computing, Communications, and Applications (TOMM)}, 18\penalty0 (2):\penalty0 1--21, 2022.

\end{thebibliography}
}
 \clearpage
\setcounter{page}{1}
\maketitlesupplementary

\renewcommand{\thesection}{\Alph{section}}
\setcounter{section}{0}

\section{Data Statistics}
\subsection{Annotations}
\begin{figure}[!htbp]
   \begin{minipage}{0.48\textwidth}
    \centering
    \includegraphics[width=\linewidth]{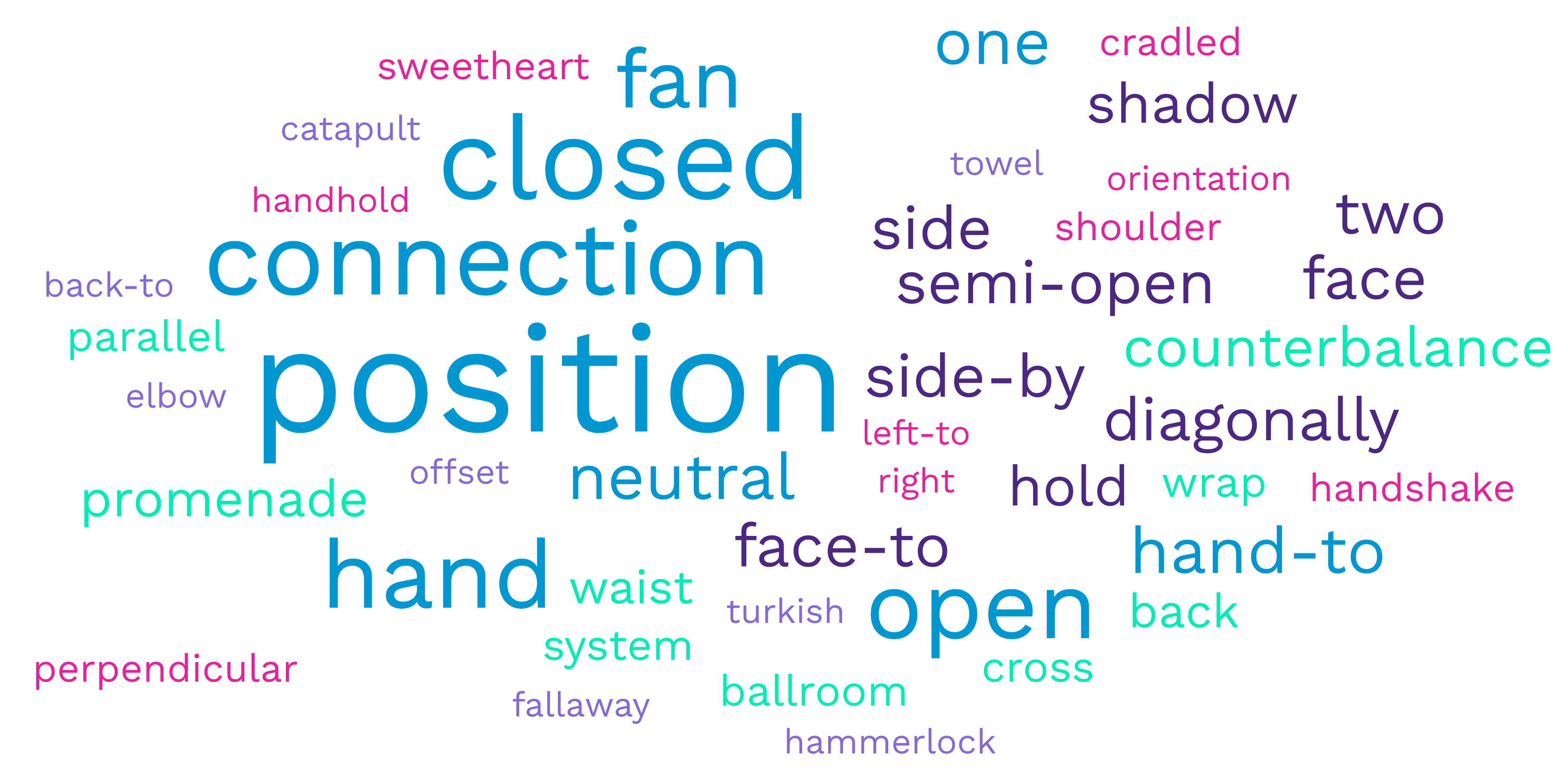}
    \caption{Spatial Relationship word cloud}
    \label{fig:word_s}
  \end{minipage}
  \hfill
  \begin{minipage}{0.48\textwidth}
    \centering
    \includegraphics[width=\linewidth]{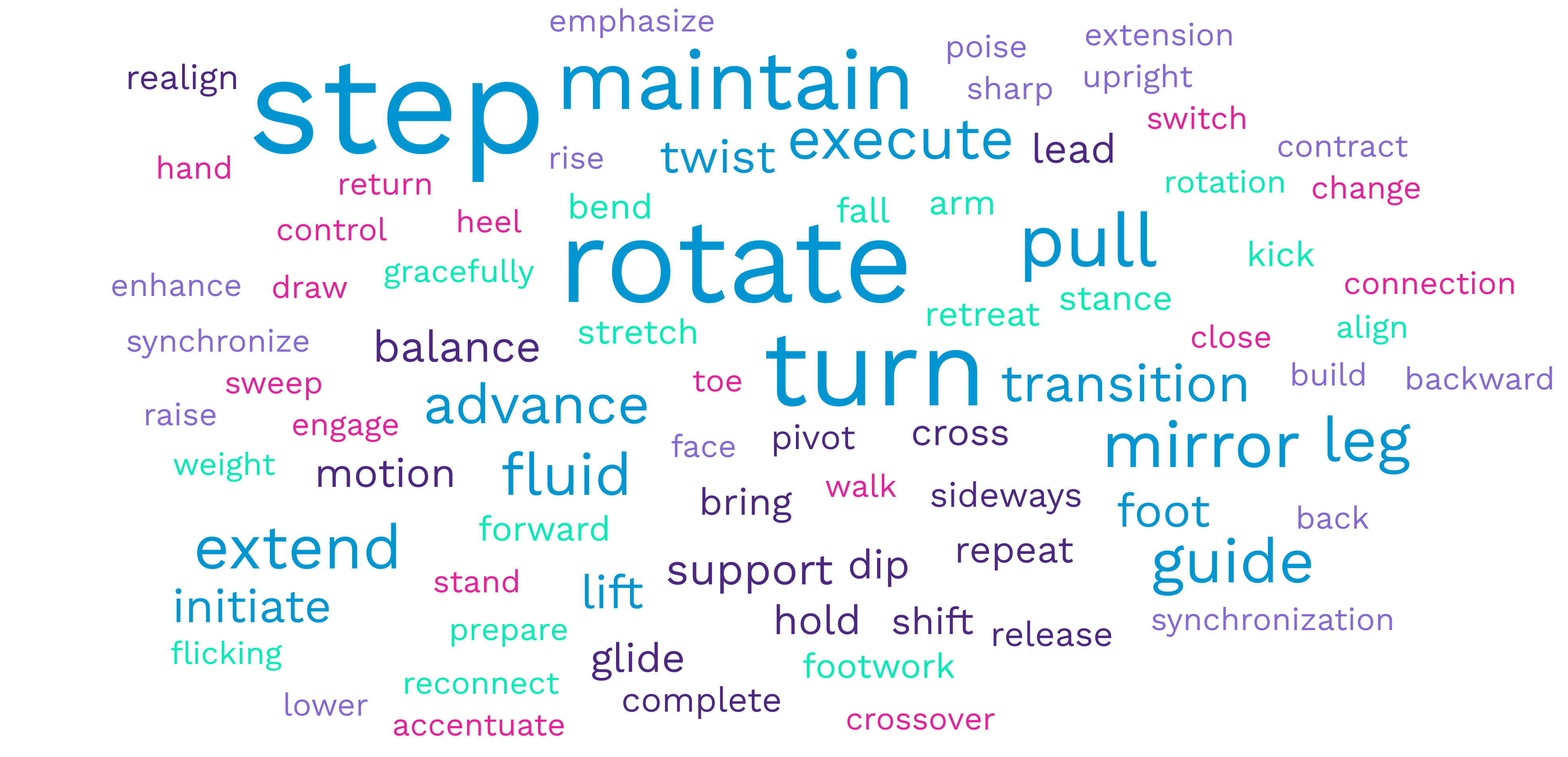}
    \caption{Body movement word cloud}
    \label{fig:word_bm}
  \end{minipage}
  \hfill
  \begin{minipage}{0.48\textwidth}
    \centering
    \includegraphics[width=\linewidth]{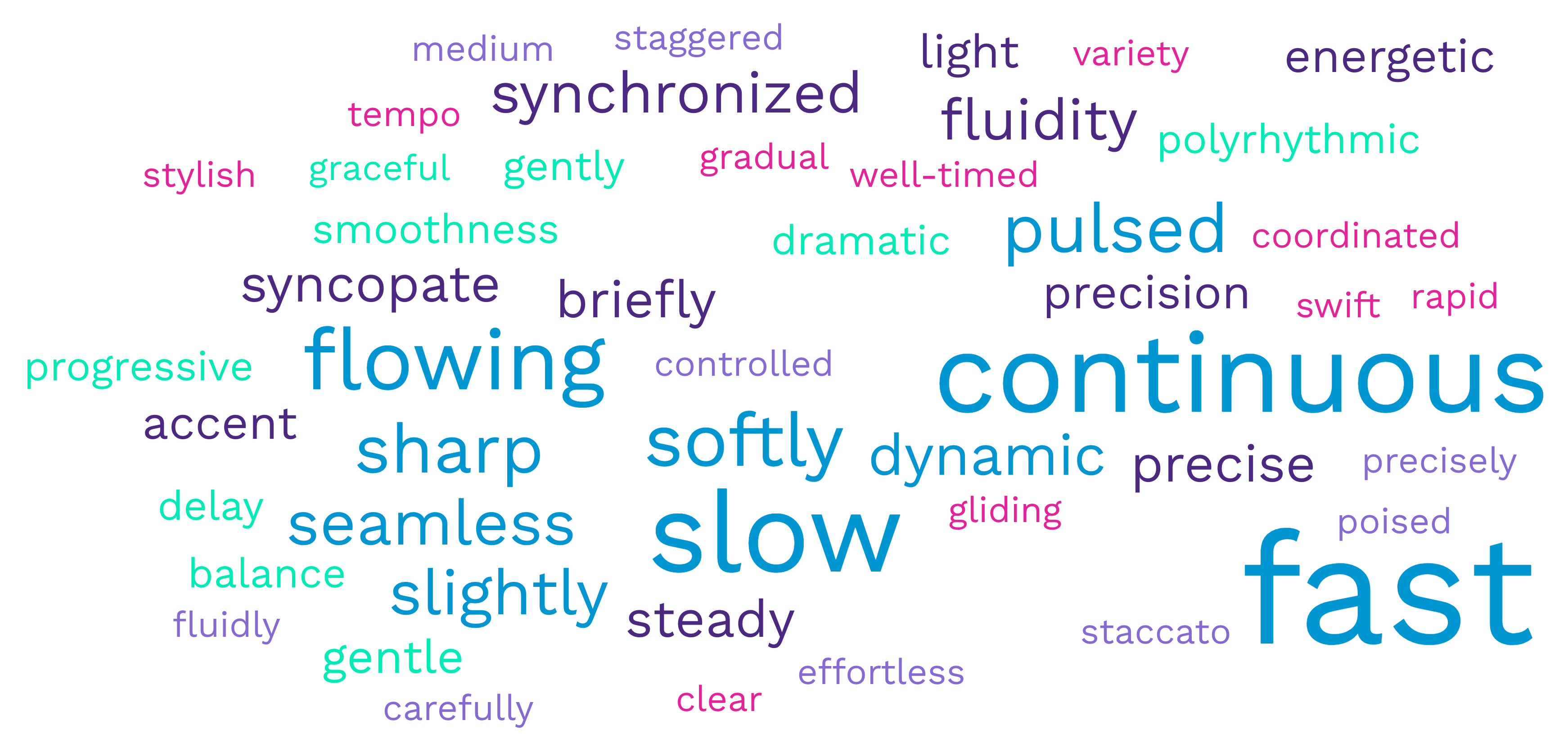}
    \caption{Rhythm word cloud}
    \label{fig:word_r}
  \end{minipage} 
\end{figure}

The annotations consist of 10,312 textual descriptions that capture detailed aspects of duet dancing. These annotations have an average length of 33.19 words, highlighting a substantial descriptive depth regarding movements, dancer interactions, and rhythmic characteristics. The total annotation corpus encompasses a vocabulary of 1,722 unique words, reflecting significant linguistic diversity. The number of annotations per genre is shown in \cref{tab:data-stats}, ranging from 548 annotations (Foxtrot) to 992 annotations (Salsa). The annotations integrate a comprehensive movement vocabulary structured around body movements (\cref{fig:word_bm}), rhythm descriptions (\cref{fig:word_r}) and spatial relationships (\cref{fig:word_s}).

\subsection{Music}
The musical characteristics of the data set also vary extensively, depending on the genre chosen. The average BPMs can be seen in the bar graph in ~\cref{fig:genre-wise}. Rumba presents the slowest music tempo, averaging 107.48 BPM, while Merengue has the fastest tempo, averaging 141.73 BPM.

\subsection{Motion Duration}
The duration statistics for the motion in each genre are shown in \cref{tab:data-stats}. $\bar{T}$ represents the average duration of a sample in the specified genre. $\hat{T}$ represents average duration of the text-annotated segments in the sample for each genre. $T$ represents the sample with the longest duration for each genre.

\section{Annotation Tool}

Fig. \ref{fig:ann_tool} shows the Interface for the Annotation website tool we developed. 
Upon logging in, users can select the preferred genre and can navigate between samples using "Previous" and "Next" buttons to select the samples they prefer annotating. The Leader dancer is color coded as blue whereas the follower is color coded as red. Users can decide how they want to optimally chunk subsamples for individual annotation based on moves executed. can sample the end time for each subsegment they want to create  To annotate, users input structured descriptions into a dedicated text box, Annotations are linked to precise time intervals, which can be set manually or through shortcut buttons like “Get Current Time” and “Go to End.” The system supports segmenting videos into discrete clips to maintain annotation precision. Annotations are first added to a "Pending Annotations" list for review before final submission, and users can edit previous annotations if necessary. The website also integrates keyboard shortcuts for efficient annotation and ensures that submitted annotations undergo validation before being finalized. This tool streamlines the annotation process while maintaining high accuracy and consistency in dance movement descriptions. The samples for annotations were divided into half among leader and follower of the sequences to be annotated.

 %specifying details such as movement terminology, spatial relationships, connection points, and rhythm.

\twocolumn[{
\vspace{-1.5em}  % Adjust the vertical space as needed
\begin{center}
  \centering
  \includegraphics[width=\textwidth]{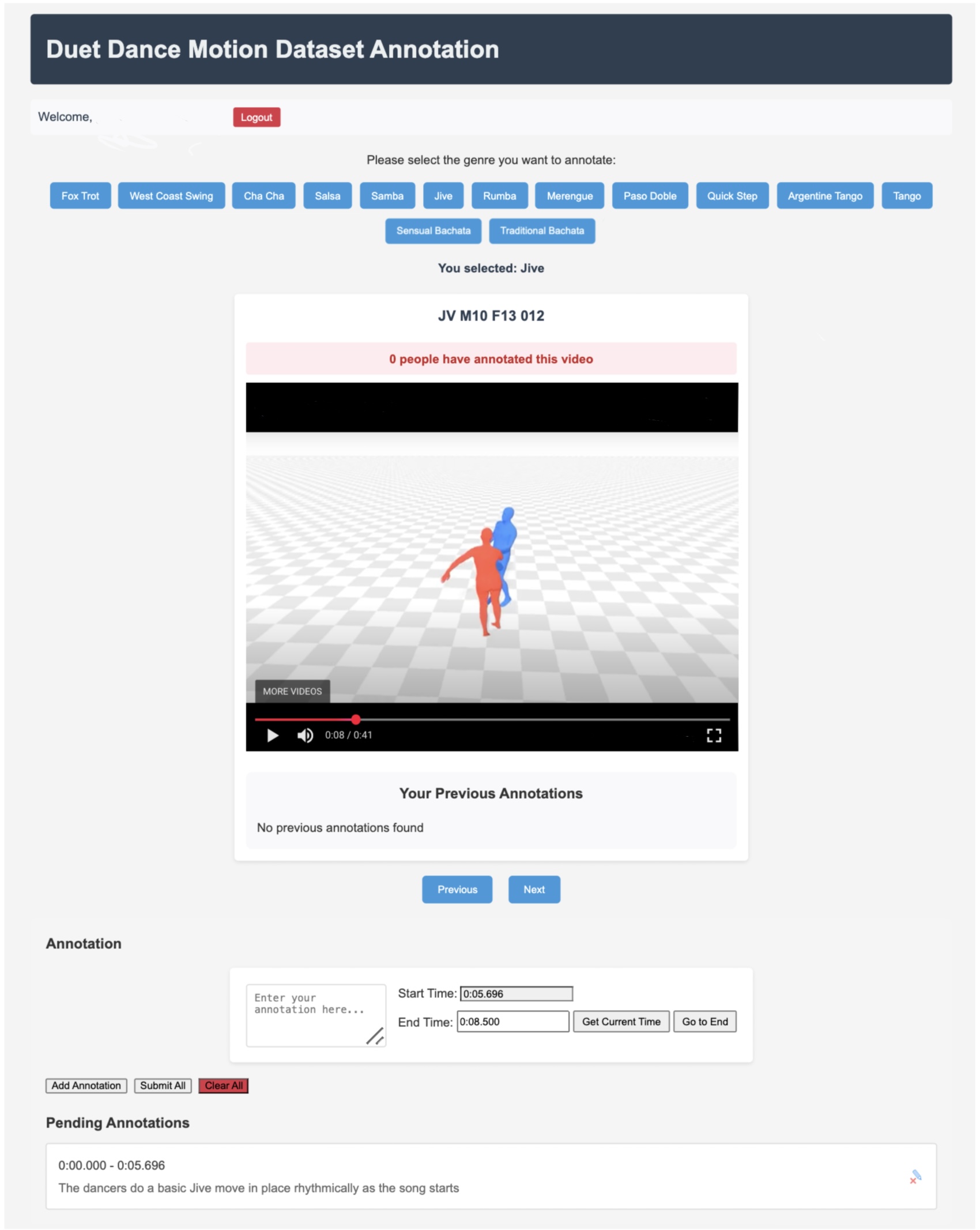}
  \captionof{figure}{Annotation Tool}
  \label{fig:ann_tool}
\end{center}
}]

\section{LLM-based Annotation Refinement}

We employed a structured prompting framework to generate refined annotations using an LLM. This process included:

\begin{enumerate}
\item \textbf{System Prompt:} The model is instructed with the following directive: \textit{“Below is a list of human-written annotations for a \{genre\} duet dance routine between a lead and a follow. As a professional dance movement analyst, refine each segment to be independently coherent, descriptive, and under 75 words. Avoid relying on prior context—each annotation must clearly describe the movement of both partners. Use the delimiter '{delimiter}' to separate annotations, and omit the original timestamps. Follow the Annotation Guidelines \{D1\}, use Duet Dance Terminology \& Vocabulary \{D2\}, and refer to Genre-specific Move Descriptions \{D3\} and provided examples \{D4\}.”}  \\ 
\item \textbf{Annotation Guidelines (D1):} A detailed set of instructions to ensure consistency between raw and refined annotations.

\item \textbf{Duet Dance Terminology (D2):} A reference document listing essential duet dance terms and structured keyword clusters organized by spatial relations, body movement, and rhythm.

\item \textbf{Genre-specific Move Descriptions (D3):} A curated compilation of genre-wise dance moves and their execution, sourced from reliable online references.

\item \textbf{Few-shot Examples (D4):} Three hand-crafted examples per genre demonstrating how raw descriptions are refined, based on expert analysis of diverse samples from the MDD dataset.

\end{enumerate}

We also use GPT-4o to extract move names from each refined annotation using the following prompt:\\

\textit{“You are a professional dance movement analyst. For each of the following refined duet dance annotations, extract a concise key dance move name mentioned or implied. If a movement corresponds to a known dance move (e.g., Underarm Turn, Hammerlock, Cross Body Lead), prefer to use that name. Use domain-specific terminology from here: \{D2\} and Genre-specific Move Descriptions\{D3\} Avoid full-sentence descriptions. Separate the move names for each annotation using a delimiter: '{delimiter}'. Do not include timestamps or any additional commentary.”}

We have included some examples for the LLM-processed descriptions and LLM-processed move names in \cref{tab:llm-text}

\begin{table*}[htbp]
\centering
\caption{Some examples showing the LLM processed text descriptions and LLM processed move names for raw text annotations belonging to different genres}
\begin{tabular}{|p{2cm}|p{6cm}|p{6cm}|p{2cm}|}
\hline
\textbf{Genre} & \textbf{Original Description} & \textbf{LLM-processed description} & \textbf{LLM-processed Move Name} \\
\hline
Jive & fallaway throwaway from closed, maintaining body contact, shifting back fro controlled release into open hold fluidly. Also, their left-to-right handhold goes through a synced transition & The leader initiates a Fallaway Throwaway from a Closed Position, maintaining body contact while shifting weight back for a controlled release, guiding the follower into an Open Hold with a fluid extension, while their left-to-right hand connection ensures a synced transition. & Fallaway Throwaway \\
\hline
Rumba & Leader faces follower in separated position facing each other, both style their arms, leader reconnects in open double handhold leading Open-to-Fan move after which they raise their arms in left-to-right handhold & With independent arm styling in a separated position while facing each other, the leader reconnects with the follower in an open double handhold, smoothly guiding her to his left executing Open-to-Fan move, after which they raise their free arms being in  left-to-right handhold & Open-to-Fan \\
\hline
Foxtrot & leader steps back from right in controlled QQ rhythm, they both go back synchronously in closed hold & The leader initiates the Back Feather, stepping backward from the right foot with a controlled yet dynamic Quick-Quick rhythm, maintaining a good flow as both partners travel backward synchronously in closed position.& Back feather \\
\hline
Waltz & both the dancers are back to back in shadow, follower in shadow. They both (follower mirrors) go into c/w rotation ending with descent & The dancers gracefully execute a Waterfall, with the follower positioned in Shadow Position, seamlessly mirroring the leader’s movement as they both ascend into an elegant clockwise rotation with a suspended, flowing descent. & Waterfall \\
\hline
 Sensual Bachata & Reverse turn: leader rotates the follower until she faces him and they both enter cross handhold & The leader initiates a Reverse turn, smoothly guiding the follower to rotate until she faces him, transitioning into a cross-handhold connection. & Reverse turn \\
\hline
\end{tabular}
\label{tab:llm-text}
\end{table*}

\section{Ablation Study}

~\cref{tab:ablation} shows the ablation results of InterGen Model for the Text-to-Duet task under different text conditions. The InterGen model trained on GPT-refined prompts outperforms both raw text and move-name versions, highlighting its effectiveness.

\section{Qualitative Visualization}
We visualize one of the samples from each genre in each category: Latin (\cref{fig:latin_genre}), Ballroom (\cref{fig:ballroom_genre}) and Social (\cref{fig:social_genre}).

\section{Generalization Experiments}

We design a set of qualitative experiments to evaluate the model’s ability to generalize across varying text prompts and dance genres. Specifically, we assess how well the model captures stylistic diversity within a genre (intra-genre generalization) and adapts to stylistic shifts across genres (inter-genre generalization), under controlled text and music variations. For each experiment, we do a paired comparison between two generated sample controlled by the experimental condition. 

We examine the following scenarios:
\begin{enumerate}
\item \textbf{Intra-genre Generalization:}
These experiments test the model’s ability to generate diverse motions within the same genre when conditioned on different textual prompts, with music held constant.

\begin{enumerate}
    \item \textbf{Same music, different text:}  
    We analyze how semantically distinct text prompts influence the resulting motions within the same genre and music context. This illustrates the model’s capacity to produce choreographic diversity while maintaining genre fidelity.

    \item \textbf{Same music, subtly different text:}  
    We evaluate the model’s sensitivity to fine-grained textual nuances—such as tempo cues, directional hints, or partner interaction specifics—demonstrating its precision in interpreting subtle prompt variations.
\end{enumerate}

\item \textbf{Inter-genre Generalization:}  
These experiments evaluate how well the model generalizes motion generation across different dance genres while controlling for text or music.

\begin{enumerate}
    \item \textbf{Same music, different text from different genres:}  
    We investigate how the model adapts to genre and prompt changes while maintaining synchronization with the same music. Here, the text prompts are sampled from different genres leading to prompting the model with move descriptions that were not present in the training data for that genre.  This highlights genre-aware style transfer and robustness to stylistic shifts.

    \item \textbf{Different music, same text:}  
    We assess how genre conditioning influences the interpretation of an identical text prompt when paired with genre-specific music. Testing this shows how the same move can be adapted to music belonging to different dance genres. This reveals how genre knowledge modulates movement vocabulary and rhythmic interpretation.
\end{enumerate}

\end{enumerate}
Examples from each scenario, generated by our adapted version of InterGen for the Text-to-Duet task, are showcased in the supplementary video. We observe that InterGen generally adheres well to the provided text prompts and music in intra-genre settings, demonstrating consistency and responsiveness to prompt variation. In inter-genre scenarios, the model exhibits a degree of stylistic adaptivity, though genre-specific fidelity may vary. Future work could focus on enhancing cross-genre generalization and enabling the synthesis of novel or unseen movement styles.

\begin{table*}[h]
  \centering % Adjust row height for better readability
  \setlength{\tabcolsep}{8pt} % Adjust column spacing for better fit
  \begin{tabular}{llccccc}
    \toprule
    \textbf{Category} & \textbf{Genre} & \textbf{Time (min)} & $\bar{T}$ & $\hat{T}$ &  $T$ & \textbf{Annotations} \\
    \midrule
    \multirow{4}{*}{Ballroom}  
      & Waltz (WZ) & 61.87 & 151.78 & 7.70 & 302.48 & 906 \\
      & Tango (TG) & 32.80 & 89.45 & 6.71 & 192.90 & 554 \\
      & Foxtrot (FT) & 31.80 & 83.68 & 9.14 & 200.91 & 548 \\
      & Quickstep (QS) & 32.30 & 69.78 & 8.64 & 190.34 & 551 \\
    \midrule
    \multirow{5}{*}{Latin}  
      & Rumba (RB) & 38.83 & 131.92 & 9.51 & 242.52 & 672 \\
      & Cha Cha (CC) & 32.82 & 99.91 & 5.08 & 260.58 & 574 \\
      & Samba (SB) & 34.20 & 90.94 & 2.28 & 149.80 & 603 \\
      & Paso Doble (PD) & 32.42 & 96.07 & 3.67 & 140.67 & 562 \\
      & Jive (JV) & 33.21 & 96.60 & 3.49 & 148.13 & 581 \\
    \midrule
    \multirow{6}{*}{Social}  
      & Salsa (SS) & 56.49 & 58.08 & 3.13 & 217.86 & 992 \\
      & Traditional Bachata (TB) & 41.38 & 119.86 &  2.96 &  221.77 & 712 \\
      & Sensual Bachata (ST) & 43.52 & 128.46 & 7.23 & 208.34 & 762 \\
      & Merengue (MR) & 36.39 & 92.36 & 3.16 &  229.73 & 642 \\
      & Argentine Tango (AT) & 61.87 & 91.03 & 6.47 & 272.92 & 634 \\
      & West Coast Swing (WC) & 50.70 & 97.40 &  3.31 & 120.14 & 894 \\
    \midrule
    \textbf{Total} & & \textbf{620.60} & & & & \textbf{10,187} \\
    \bottomrule
  \end{tabular}
  \caption{Genre-wise data distribution in the dataset, showing the total duration and the number of descriptions for each dance style.}
  \label{tab:data-stats}
\end{table*}

\begin{table*}[h]
  \centering
  \captionsetup{justification=centering}
  \resizebox{\textwidth}{!}{%
  \begin{tabular}{lccccccccc}
    \toprule
    \textbf{Methods} 
      & \multicolumn{3}{c}{\textbf{R-Precision}$\uparrow$} 
      & \textbf{FID}$\downarrow$
      & \textbf{MM Dist}$\downarrow$
      & \textbf{Diversity}$\rightarrow$
      & \textbf{MModality}$\uparrow$
      & \textbf{BED} $\uparrow$
      & \textbf{BAS}$\uparrow$ \\
    \cmidrule(lr){2-4}
    & \textbf{Top 1} & \textbf{Top 2} & \textbf{Top 3}
    & & & & & & \\
    \midrule
    Ground Truth & 0.231 & 0.3984 & 0.5219 & 0.065 & 0.0770 & 1.387 & - & 0.327 & 0.170 \\
    \midrule
    InterGen (no text, music) & 0.023 & 0.067 & 0.088 & 2.014 & 2.526 & 1.300 & \textbf{1.768} & 0.364 & 0.163 \\
    InterGen (move name, music) & 0.061 & 0.173 & 0.216 & 0.721 & 2.211 & 1.288 & 1.521 & 0.355 & \textbf{0.192} \\
    InterGen (raw text, music) & 0.911 & 0.192 & 0.244 & 0.511 & 1.722 & \textbf{1.392} & 1.411 & 0.381 & 0.190 \\
    InterGen (GPT-4o, music) & \textbf{0.105} & \textbf{0.206} & \textbf{0.302} & \textbf{0.426} & \textbf{1.532} & 1.380 & 1.352 & \textbf{0.385} & 0.185 \\
    \bottomrule
  \end{tabular}}
  \caption{Text Ablation study for InterGen (Text-to-Duet task)}
  \label{tab:ablation}
\end{table*}

\twocolumn[{
\maketitle
\vspace{-1.5em}  % Adjust the vertical space as needed
\begin{center}
  \centering
  \includegraphics[width=\textwidth]{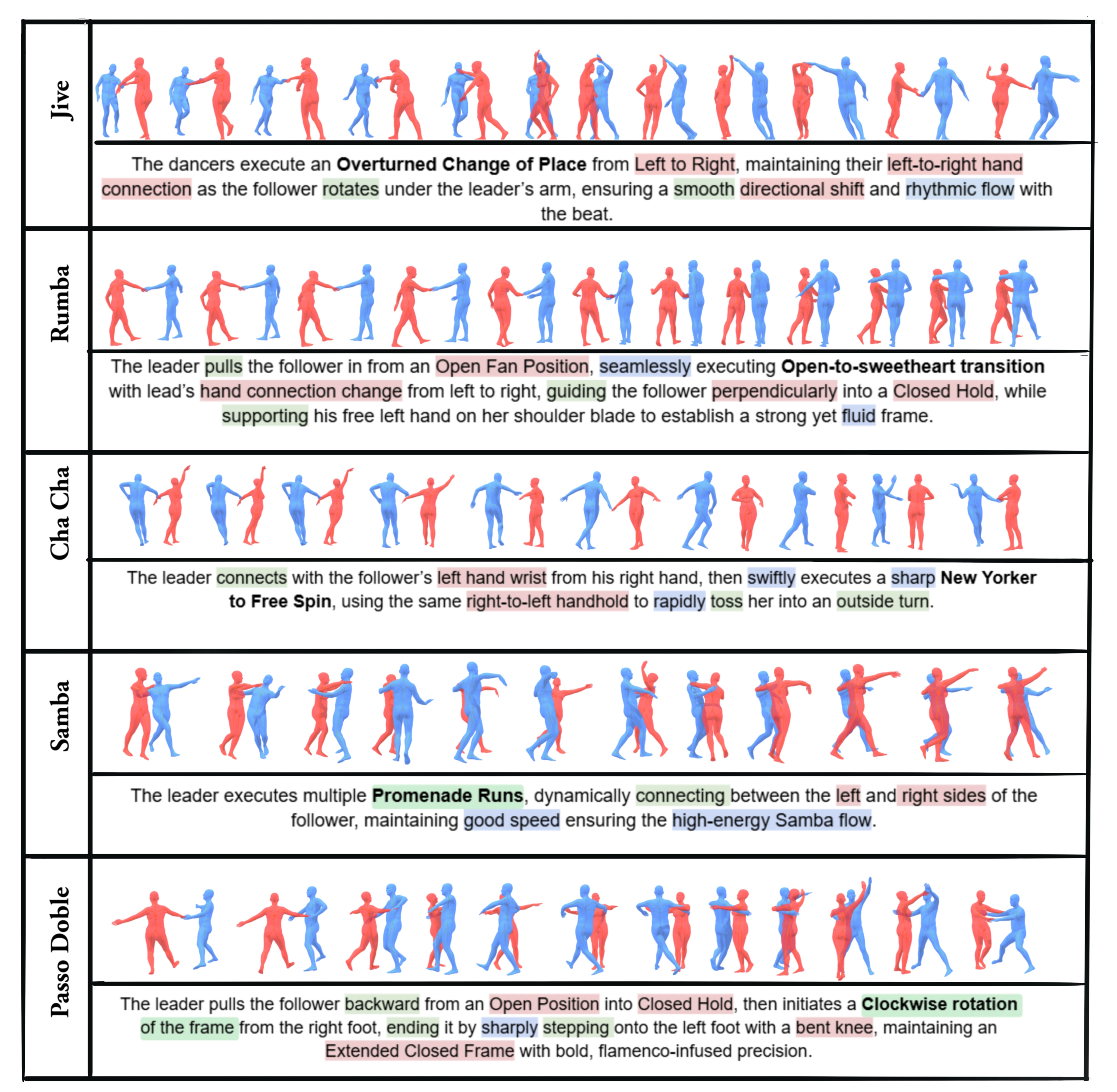}
  \captionof{figure}{Samples visualized for each Latin Category}
  \label{fig:latin_genre}
\end{center}
}]

\twocolumn[{
\maketitle
\vspace{-1.5em}  % Adjust the vertical space as needed
\begin{center}
  \centering
  \includegraphics[width=\textwidth]{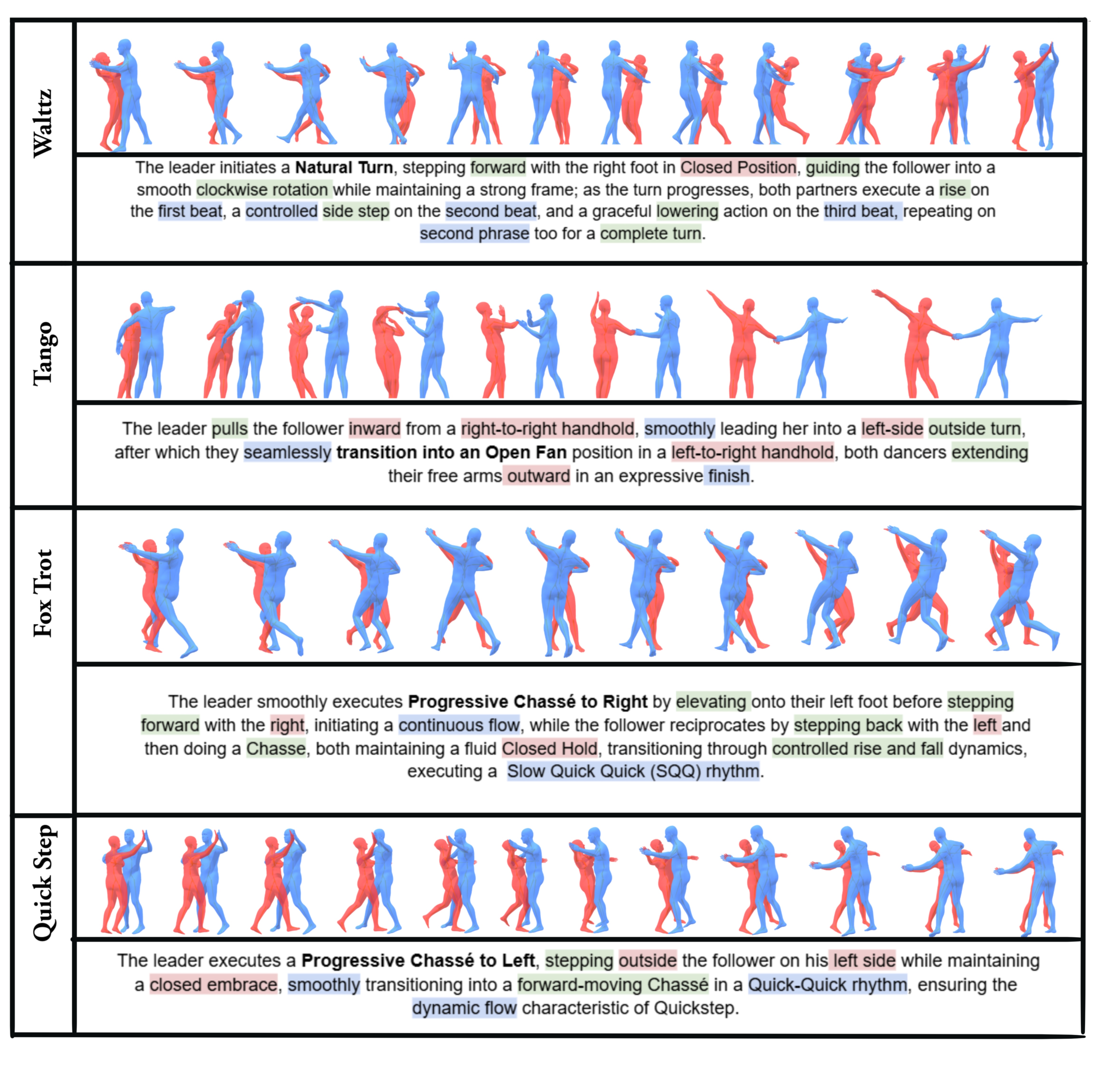}
  \captionof{figure}{Samples visualized for each Ballroom Category}
  \label{fig:ballroom_genre}
\end{center}
}]

\twocolumn[{
\maketitle
\vspace{-1.5em}  % Adjust the vertical space as needed
\begin{center}
  \centering
  \includegraphics[width=\textwidth]{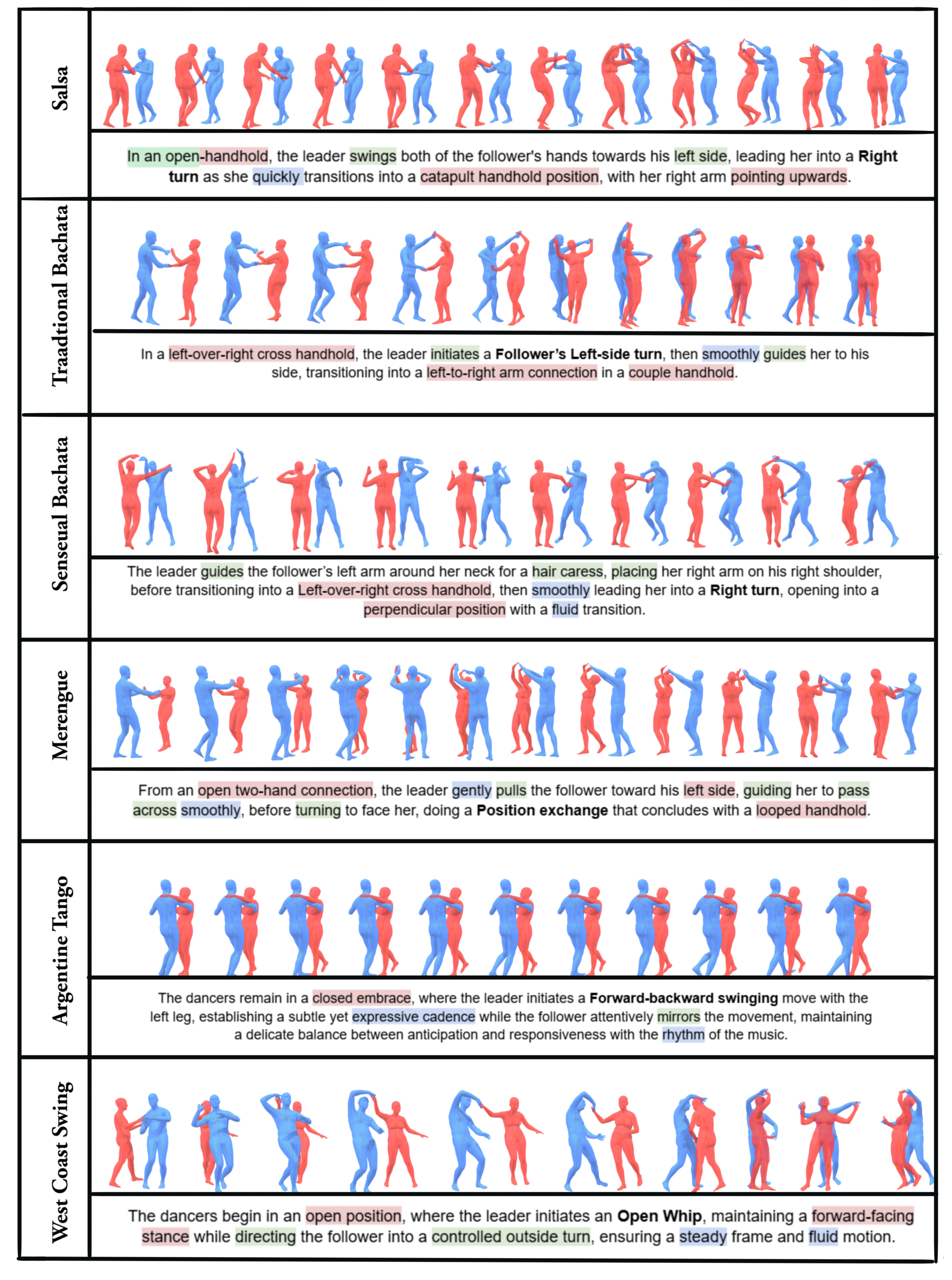}
  \captionof{figure}{Samples visualized for each Social Category}
  \label{fig:social_genre}
\end{center}
}]

% {
%     \small
%     \bibliographystyle{ieeenat_fullname}
%     \bibliography{main}
% }

\end{document}